\pdfoutput=1
\documentclass[11pt, a4paper]{gdm/googledeepmind}


\pdftrailerid{redacted}

\makeatletter
\renewcommand\bibentry[1]{\nocite{#1}{\frenchspacing\@nameuse{BR@r@#1\@extra@b@citeb}}}
\makeatother

\usepackage[authoryear,sort&compress,round]{natbib}
\usepackage[colorlinks=true,allcolors=blue]{hyperref}
\hypersetup{citecolor=[HTML]{2420cf}, linkcolor=[HTML]{2420cf}, urlcolor=[HTML]{2420cf}}
\usepackage{url}
\usepackage{xurl}

\usepackage{dsfont}
\usepackage{gdm/gdm-colors}
\usepackage{graphicx}
\usepackage{wrapfig}
\usepackage{capt-of}
\usepackage{booktabs}
\usepackage{nicefrac}
\usepackage{microtype}
\usepackage{enumitem}
\usepackage{algorithm}
\usepackage{algpseudocode}
\usepackage{pifont}
\usepackage{marvosym}
\usepackage{fontawesome5}
\usepackage{multirow}
\usepackage{makecell}
\usepackage{stfloats}
\usepackage{listings}
\usepackage{mdframed}
\usepackage{soul}
\usepackage{array}
\usepackage{amssymb}
\usepackage{mathtools}
\usepackage{amsthm}
\usepackage{amsmath}
\usepackage{bm}

\graphicspath{{figures/}}

\definecolor{hlcol}{HTML}{FFF4CC}
\sethlcolor{hlcol}
\newcolumntype{g}{>{\columncolor{gray!15}}c}

\lstdefinestyle{promptstyle}{
  backgroundcolor=\color{gray!6},
  frame=single,
  rulecolor=\color{gray!45},
  basicstyle=\fontsize{7.0pt}{8.5pt}\selectfont\ttfamily,
  breaklines=true,
  breakatwhitespace=false,
  columns=fullflexible,
  keepspaces=true,
  showstringspaces=false,
  xleftmargin=6pt,
  xrightmargin=6pt,
  aboveskip=4pt,
  belowskip=4pt,
  captionpos=b,
}
\mdfdefinestyle{examplebox}{
  backgroundcolor=gray!5,
  linecolor=gray!50,
  linewidth=0.6pt,
  innerleftmargin=7pt,
  innerrightmargin=7pt,
  innertopmargin=5pt,
  innerbottommargin=5pt,
  skipabove=4pt,
  skipbelow=4pt,
}

\newcommand{\DTTightSection}{}
\newcommand{\DTTightSubsection}{}
\newcommand{\DTTightSubsubsection}{}
\newcommand{\DTTightParagraph}{}
\newcommand{\DTTightBetween}{}
\newcommand{\DTTightList}{}
\newcommand{\DTTightFloatTop}{}
\newcommand{\DTTightCaption}{}
\newcommand{\DTTightTableCaption}{}
\newcommand{\DTTightFloatBottom}{}

\renewcommand{\arraystretch}{1.12}
\renewcommand{\eqref}[1]{Eq.~\ref{#1}}

\theoremstyle{plain}

\theoremstyle{definition}

\theoremstyle{remark}

\newcommand{\DTAppendixTOCSection}[2]{%
  \noindent\hyperref[#1]{\textbf{\ref*{#1}\quad #2}}\dotfill\pageref{#1}\par%
}
\newcommand{\DTAppendixTOCSubsection}[2]{%
  \noindent\hspace*{1.8em}\hyperref[#1]{\ref*{#1}\quad #2}\dotfill\pageref{#1}\par%
}
\newcommand{\DTAppendixContentsPage}{%
  \clearpage
  \phantomsection
  \pdfbookmark[1]{Appendices}{appendices}
  \section*{Appendices}
  \vspace{0.35em}
  \begingroup
  \footnotesize
  \setlength{\parskip}{1.3pt}
  \DTAppendixTOCSection{app:related}{Extended Related Work}
  \DTAppendixTOCSubsection{app:related:agents}{Tool-Augmented LLM Agents}
  \DTAppendixTOCSubsection{app:related:memory}{LLM Personalization and Memory}
  \DTAppendixTOCSubsection{app:related:kt}{Knowledge Tracing and Intelligent Tutoring Systems}
  \DTAppendixTOCSubsection{app:related:edu}{LLM-Based Educational Agents}
  \DTAppendixTOCSection{app:prompts}{Implementation Details of DeepTutor System}
  \DTAppendixTOCSubsection{app:algorithm}{Closed-Loop Personalized Tutoring}
  \DTAppendixTOCSubsection{app:tutorbot}{TutorBot Autonomous Agent Loop}
  \DTAppendixTOCSection{app:tutorbench}{TutorBench Construction Details}
  \DTAppendixTOCSubsection{app:tutorbench:sources}{PDF Source Inventory}
  \DTAppendixTOCSubsection{app:tutorbench:task}{Task Generation and Rejection Sampling}
  \DTAppendixTOCSubsection{app:tutorbench:traceforest}{Trace Forest and Benchmark Runtime Memory}
  \DTAppendixTOCSection{app:eval}{Evaluation Protocol and Hyperparameters}
  \DTAppendixTOCSubsection{app:eval:interactive}{First-Person Interactive Evaluation Details}
  \DTAppendixTOCSubsection{app:results:breakdown}{Cross-Domain Metric Table}
  \DTAppendixTOCSubsection{app:eval:baselines}{Baseline and Variant Definitions}
  \DTAppendixTOCSubsection{app:eval:human_alignment}{Human Preference Alignment Study}
  \DTAppendixTOCSubsection{app:eval:general}{General Problem-Solving Setup}
  \DTAppendixTOCSubsection{app:eval:hparams}{Agent Hyperparameters}
  \DTAppendixTOCSubsection{app:eval:impl}{Implementation Details}
  \DTAppendixTOCSubsection{app:eval:backbones}{Backbone Models}
  \endgroup
  \clearpage
}

\titleformat{\paragraph}[runin]{\bfseries}{}{0pt}{#1}
\titlespacing*{\paragraph}{0pt}{0.8ex plus 0.3ex minus 0.2ex}{0.8em}

\title{DeepTutor: Towards Agentic Personalized Tutoring}
\correspondingauthor{chaohuang75@gmail.com, bingxizhao39@gmail.com}
\renewcommand{\today}{2026-05-07}

\author{%
  {\fontsize{10.5}{13}\selectfont\bfseries Bingxi Zhao\textsuperscript{*}, Jiahao Zhang\textsuperscript{*}, Xubin Ren, Zirui Guo, Tianzhe Chu, Yi Ma, Chao Huang\textsuperscript{\textdagger}}\par\vspace{3pt}
  {\normalfont\bfseries\itshape\fontsize{10}{12}\selectfont The University of Hong Kong}\par\vspace{4pt}
  {\normalfont\small\href{https://github.com/HKUDS/DeepTutor}{\textcolor{black}{\faIcon{github}\ https://github.com/HKUDS/DeepTutor}}}%
}

\makeatletter
\fancypagestyle{firststyle}{
  \fancyhf{}
  \fancyhead[L]{\includegraphics[width=120pt]{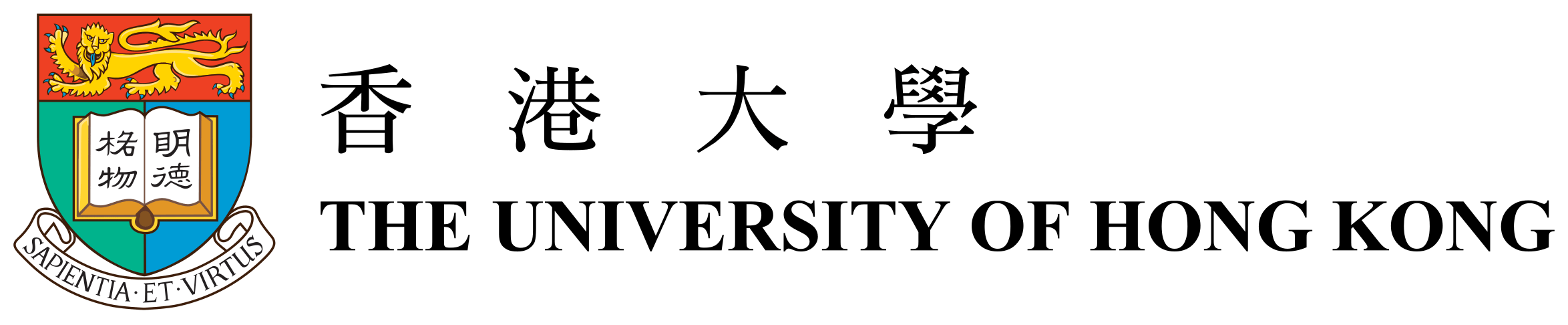}}
  \fancyhead[C]{}
  \fancyhead[R]{\fontsize{10.5}{13}\selectfont Technical Report.}
  \fancyfoot[L]{\footerfont\textsuperscript{*}Equally contributed.\quad\textsuperscript{\textdagger}Corresponding author.\\ \itshape Contact: \the\correspondingauthor \\}
  \fancyfoot[C]{\footerfont\bfseries\relax}
  \fancyfoot[R]{}

}
\makeatother

\begin{abstract}
Education is one of the most promising real-world applications for Large Language Models (LLMs).
However, current LLMs rely on static pre-training knowledge and lack adaptation to individual learners,
while existing RAG systems fall short in delivering personalized, guided feedback.
To bridge this gap, we present DeepTutor, a fully open-source agentic framework that unifies citation-grounded problem tutoring with difficulty-calibrated question generation.
A hybrid personalization engine couples static knowledge grounding with dynamic learner memory,
continuously adapting each interaction to the student's evolving needs.
The same personalization substrate further extends to adaptive learning workflows, interactive books, and proactive multi-channel tutoring agents.
To evaluate personalized tutoring, we introduce TutorBench, an interactive benchmark incorporating
customized learner profiles grounded in university-level curricula across five domains.
We further propose an LLM-based first-person interactive evaluation protocol that conducts
assessments via a profile-driven student simulator.
Complementary evaluations on established benchmarks, supported by human-alignment and ablation studies, confirm the framework's robustness and general utility.
Results show that DeepTutor improves personalized metrics by 10.8\% on average and strengthens general agentic reasoning across five backbone models by 29.4\%.
\end{abstract}

\begin{document}

\maketitle

\section{Introduction}
\DTTightSection

Education represents one of the most impactful and rapidly evolving real-world domains for the deployment of large language models.
An effective human tutor does far more than supply correct answers. They diagnose misconceptions, scaffold explanations to the learner's level, and continuously pose targeted practice that turns weaknesses into strengths~\citep{letourneau2025systematicreviewaidriven}.
While LLMs exhibit impressive fluency in educational dialogues, replicating this tightly integrated set of pedagogical skills within a unified, adaptive system remains an open challenge~\citep{chu-etal-2025-llm}.
\DTTightBetween

\begin{figure}[t!]
    \centering
    \DTTightFloatTop
    \includegraphics[width=\columnwidth]{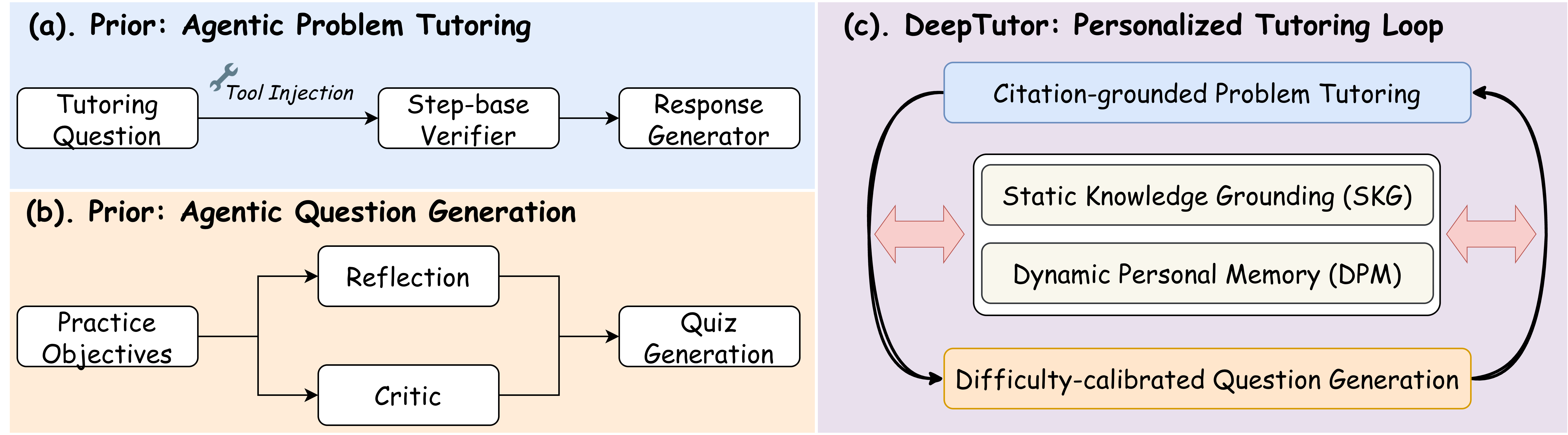}
    \DTTightCaption
    \caption{From isolated education workflows to closed-loop personalized tutoring.
    (a,b) Prior tutoring and question-generation agents operate task-locally; (c) DeepTutor uses a Hybrid Personalization Engine, combining Static Knowledge Grounding (SKG) and Dynamic Personal Memory (DPM), to form a closed loop between tutoring traces and personalized practice.}
    \label{fig:compare}
    \DTTightFloatBottom
\end{figure}

Existing approaches tackle fragments of this challenge in isolation, as illustrated in Figure~\ref{fig:compare}.
On the problem-tutoring side, recent LLM tutoring work has moved beyond direct answer generation by evaluating open-ended pedagogical behavior, diagnosing student reasoning errors, and aligning models toward guided problem solving~\citep{daheim2024stepwise,huang2025mathtutorbench,dinucujianu2025pedagogy}.
This makes responses more targeted, but the resulting diagnosis is usually consumed within the same tutoring turn.
On the question-generation side, LLM-guided generation methods use planning or self-refinement to control question content, difficulty, and educational objectives~\citep{li2024planning,yao2025mcqg,cheng2025objectives}.
These systems insightfully improve item quality, yet their generation process is typically driven by predefined objectives rather than the student's recent tutoring traces.
Together, these workflows remain task-local: tutoring traces do not condition the next practice item, and practice outcomes do not update future explanations.
Both shortcomings share a root cause: current systems lack a fine-grained, evolving model of the learner. At best, they track coarse skill inventories rather than the reasoning traces that reveal \textit{how} a student errs~\citep{park2024empoweringpersonalizedlearning,wang2025llmpoweredmultiagentframeworkgoaloriented,corbett1994knowledge,piech2015deep,zhang2017dkvmn,ghosh2020akt}.
\DTTightBetween

Compounding this system gap is an \textit{evaluation} gap.
Most educational benchmarks adopt an instructor-centric perspective, testing whether an LLM follows sound pedagogical principles or produces factually correct answers, while treating the student as a generic receiver~\citep{chu-etal-2025-llm,kurdi2020systematic}.
The few learner-aware efforts represent students with coarse proficiency labels that are disconnected from any specific course material.
As a result, whether a tutoring system can truly \textit{adapt} to an individual learner across a multi-turn conversation has remained largely untested.
\DTTightBetween

We argue that addressing both gaps requires a \textit{closed interactive loop}: weaknesses exposed during problem tutoring should directly shape which questions are generated next, while the learner's performance on those questions should in turn refine the model that informs future explanations~\citep{black1998assessment,shute2008focus}.
Realizing such a loop demands two ingredients: a structured memory that captures fine-grained interaction histories rather than scalar skill scores, and a unified system whose modules share and continuously update that memory.
\DTTightBetween

To bridge these gaps, we propose \textbf{DeepTutor}, which unifies problem tutoring and question generation through a shared personalization engine.
Both pipelines are coupled through a \textit{trace forest}, a hierarchical memory whose specialized agents continuously distill interaction traces into an evolving learner profile, creating a closed tutoring cycle in which every interaction personalizes the next.
To enable personalized evaluation, we also construct \textbf{TutorBench}, a student-centric benchmark built from university-level materials across five broad disciplines. Each entry couples a source-grounded learner profile with diagnosed knowledge gaps and an interactive tutoring task, and a first-person student simulator drives multi-turn dialogue that tests adaptive behavior end to end.
\DTTightBetween

Our main contributions are as follows:
\DTTightList
\begin{itemize}[nosep,leftmargin=*]
  \item We introduce DeepTutor, an open-source agentic tutoring framework that closes the loop between citation-grounded problem tutoring and difficulty-calibrated practice generation through shared learner context.
  \item We design a hybrid personalization engine that combines Static Knowledge Grounding with Dynamic Personal Memory, using a trace forest to distill multi-turn interactions into fine-grained, reusable learner profiles.
  \item We characterize how the same personalization substrate extends beyond reactive tutoring to broader adaptive learning workflows, interactive books, and proactive multi-channel tutoring agents, while separating these system extensions from the quantitatively evaluated tutoring core.
  \item We construct TutorBench, a student-centric benchmark with first-person interactive evaluation across five university-level disciplines, and show that DeepTutor improves personalized tutoring quality while generalizing across domains and backbone models.
\end{itemize}
\DTTightList

\section{DeepTutor Framework}
\label{sec:method}
\DTTightSection

\begin{figure*}[t]
    \DTTightFloatTop
    \centering
    \includegraphics[width=\textwidth]{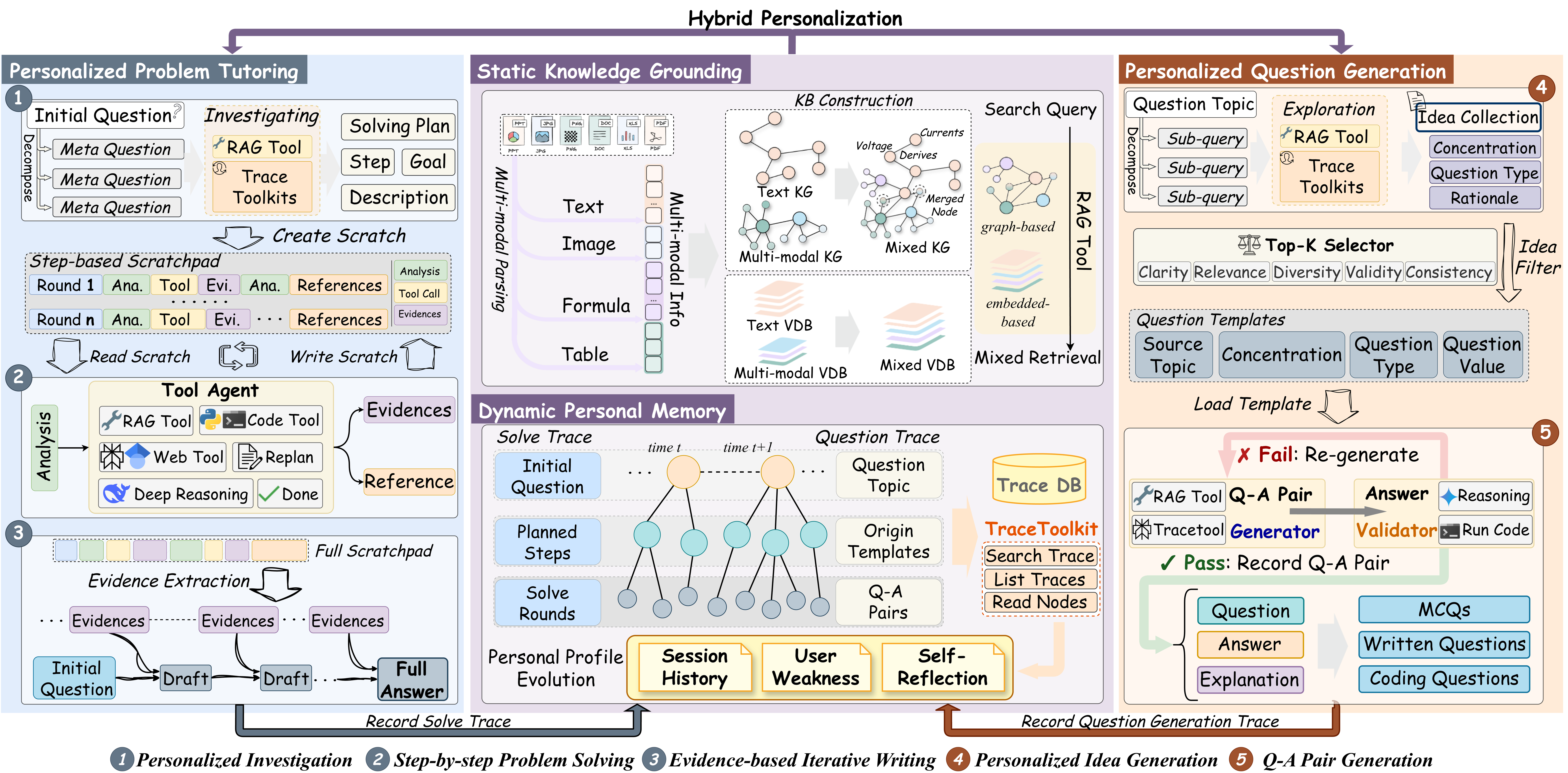}
    \DTTightCaption
    \caption{Overview of DeepTutor.
    Static Knowledge Grounding supplies course context, Dynamic Personal Memory supplies learner context, and the Hybrid Personalization layer feeds both problem tutoring and question generation.
    Completed interactions update the trace forest and learner profile, forming a closed tutoring cycle.}
    \label{fig:overview}
    \DTTightFloatBottom
\end{figure*}

As shown in Figure~\ref{fig:overview}, DeepTutor unifies two tutoring tasks through a shared Hybrid Personalization Engine: \emph{problem tutoring}, which produces citation-grounded, learner-calibrated guidance $a$ for a student question $q$, and \emph{question generation}, which produces customized practice items for a topic $\tau$ targeting the learner's gaps.
The engine couples course knowledge $\mathcal{K}$ with an evolving learner profile $\mathcal{D} = (\mathcal{D}_s, \mathcal{D}_w, \mathcal{D}_r)$, assembling a personalization context $\mathcal{C}_{\text{mem}}$ before each agent step.
Problem tutoring then proceeds through investigation, guided solving, and iterative writing (\S\ref{sec:solve}), while question generation separates idea selection from verified question--answer--explanation construction (\S\ref{sec:generate}).
Each completed interaction updates the profile and trace history, forming the closed loop summarized in Algorithm~\ref{alg:loop}.
\DTTightBetween

% ===========================================================================
\subsection{Hybrid Personalization Engine}
\label{sec:personalization}
\DTTightSubsection
% ===========================================================================

Both pipelines require two complementary forms of context: \emph{domain expertise} and \emph{learner awareness}.
DeepTutor provides the former through Static Knowledge Grounding and the latter through Dynamic Personal Memory, unified by a profile injection mechanism that assembles role-specific context for each agent.

\subsubsection{Static Knowledge Grounding (SKG)}
\label{sec:rag}
\DTTightSubsubsection
To support real course materials with heterogeneous content, we decompose source documents into atomic content units and preserve modality-aware structure when available~\citep{wang2024mineruopensourcesolutionprecise}.
These units are then indexed through two complementary structures. A \emph{knowledge graph} $\mathcal{G}$ organizes structural and contextual relations across the extracted content units, while dense encoders project all units into an \emph{embedding index} $\mathcal{B}$~\citep{guo2025raganythingallinoneragframework}.
Together, $\mathcal{G}$ and $\mathcal{B}$ form a dual-index retrieval substrate over $\mathcal{K}$.
\DTTightBetween

At query time, graph traversal over $\mathcal{G}$ captures explicit relationships while dense search over $\mathcal{B}$ surfaces semantically similar content. The two candidate sets are fused via reciprocal rank fusion~\citep{cormack2009reciprocal}, deduplicated, and truncated to a context budget, yielding the domain grounding $\mathcal{C}_{\text{rag}}$.
\DTTightBetween

% Real-world course materials are inherently multimodal: textbooks interleave prose with figures, equations, and data tables, each carrying domain knowledge that plain-text flattening would distort.
% We therefore decompose source documents into atomic content units, each tagged by modality and processed by a dedicated extractor that preserves its internal structure, including captions for figures, anchored definitions for equations, header-cell relationships for tables~\cite{wang2024mineruopensourcesolutionprecise}.

% These units are indexed through two complementary structures.
% A \emph{knowledge graph} $\mathcal{G}$ merges a cross-modal subgraph that links non-textual elements to their contextual neighborhood with a text-based semantic subgraph.
% In parallel, dense encoders project all units into a combined \emph{embedding index} $\mathcal{B}$.
% Together, $\mathcal{G}$ and $\mathcal{B}$ form a dual-index retrieval substrate over $\mathcal{K}$.

% At query time, graph traversal over $\mathcal{G}$ captures explicit relationships and multi-hop reasoning chains, while dense search over $\mathcal{B}$ surfaces semantically similar content beyond the graph's structural reach.
% The two candidate sets are fused via reciprocal rank fusion~\citep{cormack2009reciprocal}, deduplicated, and truncated to a context budget, yielding the domain grounding $\mathcal{C}_{\text{rag}}$.

\subsubsection{Dynamic Personal Memory (DPM)}
\label{sec:memory}
\DTTightSubsubsection

While the knowledge base captures \emph{what the course teaches}, the dynamic memory captures \emph{how the individual learner has engaged with it}.
Central to this component is the \textit{Trace Forest} $\mathcal{F}$, where each tree records one complete tutoring interaction as a multi-resolution, semantically searchable artifact.
Nodes are organized into three levels: \textit{Level~1} stores session-level input and a global summary; \textit{Level~2} captures intermediate planning units from task decomposition; and \textit{Level~3} preserves fine-grained execution records including tool outputs, evidence, and validation outcomes.
Every node carries a dense embedding, enabling similarity-based retrieval across the entire forest.
This design is inspired by recent agent memory systems that externalize long-term context and organize experience hierarchically across reusable workflows or interaction traces~\citep{packer2023memgpt,wang2024agentworkflowmemory,qin2025himem,yang2025gmemory}.
The concrete node types at each level are task-specific and detailed alongside the respective pipelines (\S\ref{sec:solve}--\S\ref{sec:generate}).
\DTTightBetween

Rather than treating $\mathcal{F}$ as a passive store, we expose it through a programmatic \emph{TraceToolkit} with three operations: \textsc{SearchTrace} for semantic ANN retrieval across $\mathcal{F}$, \textsc{ListTraces} for filtered enumeration by time, task type, or topic, and \textsc{ReadNodes} for full-content retrieval together with ancestor paths. This toolkit enables every agent to explore the learner's history at the resolution it needs.
\DTTightBetween

\DTTightParagraph
\paragraph{Profile Construction.}
\label{sec:injection}
Trace Forest provides an analyzable evidence space, but learner profiles are not produced by passively summarizing the latest interaction.
For each new trace, three specialized memory agents actively query the TraceToolkit to retrieve related prior sessions, inspect fine-grained nodes when needed, and compare the latest behavior against cross-session patterns before updating $\mathcal{D}=(\mathcal{D}_s,\mathcal{D}_w,\mathcal{D}_r)$.
The resulting profile contains three complementary views: $\mathcal{D}_s$ summarizes session history such as topics covered, solving paths, and performance trends; $\mathcal{D}_w$ maintains an evidence-backed inventory of recurring confusions, wrong-answer patterns, and active or resolved knowledge gaps; and $\mathcal{D}_r$ records pedagogical self-reflections that guide future interactions.
This process makes profile construction a tool-mediated analysis procedure rather than a single-pass textual summary, grounding personalization in observable trace evidence rather than coarse latent mastery scores.
\DTTightBetween

Before each agent step, the system assembles a personalization context $\mathcal{C}_{\text{mem}}$ through two channels:
(i)~\emph{active trace retrieval}, where the TraceToolkit surfaces the most relevant nodes with budget allocated proportionally across levels to balance broad session context with fine-grained precedents;
and (ii)~\emph{role-specific profile excerpting}, which routes different slices of $\mathcal{D}$ to different agents based on their function (e.g., the planner receives $\mathcal{D}_s$ and $\mathcal{D}_w$; the writer receives $\mathcal{D}_r$; generation agents receive $\mathcal{D}_w$ alongside historical question patterns), reducing irrelevant context for each role.
The total token budget is dynamically partitioned between $\mathcal{C}_{\text{rag}}$ and $\mathcal{C}_{\text{mem}}$.
\DTTightBetween

% ===========================================================================
\subsection{Personalized Problem Tutoring}
\label{sec:solve}
\DTTightSubsection
% ===========================================================================

Prior tool-augmented agents~\citep{yao2023reactsynergizingreasoningacting,wu-etal-2025-agentic,li2025intheflowagenticoptimizationeffective,chen2026iterresearch} fold planning, retrieval, and composition into a single reasoning loop.
This can suffice for short factoid queries but can become brittle in tutoring, where thorough investigation and personalized presentation compete for the same context window.
Inspired by \textit{scaffolding theory}, DeepTutor disentangles problem tutoring into three stages: investigate first, guide the solution incrementally, then write adaptively~\citep{wood1976role}. These steps correspond to \ding{172}--\ding{174} in Appendix Algorithm~\ref{alg:loop}.
\DTTightBetween

\DTTightParagraph
\paragraph{Stage \ding{172}: Personalized Investigation.}
The planner first runs a lightweight investigation pass before
committing to a plan: it decomposes $q$ into meta-questions and gathers
evidence across $\mathcal{K}$ (via RAG), the trace forest (via the
TraceToolkit), and auxiliary tools when needed.
Conditioning jointly on the resulting domain context $\mathcal{C}_{\text{rag}}$ and learner context $\mathcal{C}_{\text{mem}}$, it produces a tutoring plan $\mathcal{P} = \langle s_1, \ldots, s_K \rangle$ of concrete, annotated sub-goals.
This \emph{investigate-before-plan} design yields sub-goals that are specific to the learner's actual gaps rather than generically phrased (e.g., ``review chain rule in trigonometry'' instead of a vague ``review calculus'').
\DTTightBetween

\DTTightParagraph
\paragraph{Stage \ding{173}: Step-by-step Guided Solving.}
For each sub-goal $s_k$, the solver first analyzes the target, then acts and records notes using a shared tool suite~\citep{yao2023reactsynergizingreasoningacting}, conditioned on both $\mathcal{C}_{\text{rag}}$ and $\mathcal{C}_{\text{mem}}$.
Two mechanisms manage the growing reasoning context: \emph{self-notes} distill each step's outcome into a concise takeaway that later steps reference in lieu of verbose intermediates, and \emph{hierarchical compression} progressively summarizes completed sub-goals into compact digests, freeing context capacity for deeper reasoning on later ones.
When the current plan proves inadequate, \emph{adaptive replanning} revises remaining sub-goals while preserving completed work.
\DTTightBetween

\DTTightParagraph
\paragraph{Stage \ding{174}: Evidence-based Iterative Writing.}
Rather than concatenating raw tool outputs, the writer extracts structured evidence from the scratchpad and constructs the answer through successive refinement passes, reconciling potentially conflicting findings.
The learner context $\mathcal{C}_{\text{mem}}$ steers both the depth and tone of the explanation to the learner's \textit{Zone of Proximal Development}~\citep{vygotsky1978mind}: \textit{beginners} receive scaffolded step-by-step derivations, while \textit{proficient learners} receive concise summaries foregrounding key insights.
Every externally grounded factual claim carries a traceable citation to
retrieved evidence, typically from $\mathcal{K}$ and, when enabled,
auxiliary web sources.
\DTTightBetween

% ===========================================================================
\subsection{Personalized Question Generation}
\label{sec:generate}
\DTTightSubsection
% ===========================================================================

A tutor must also pose the right questions: ones that are meaningful, factually grounded, and calibrated to the individual learner.
We propose a two-stage architecture that separates \emph{what to ask} from \emph{how to ask and verify}; these stages correspond to \ding{175}--\ding{176} in Appendix Algorithm~\ref{alg:loop}.
\DTTightBetween

\DTTightParagraph
\paragraph{Stage \ding{175}: Personalized Idea Generation.}
Rather than generating questions directly from a bare topic, the idea agent first maps the conceptual landscape around $\tau$ through the lens of the individual learner.
Conditioning jointly on $\mathcal{C}_{\text{rag}}$ and $\mathcal{C}_{\text{mem}}$, including past mistakes, difficulty trends, and related problems surfaced via the TraceToolkit, the idea agent produces a pool of candidate ideas, each specifying a target concept, question format (MCQ, written-response, or coding), and a personalized rationale grounded in diagnosed gaps.
An evaluator-driven filtering and ranking procedure iteratively prunes
ill-formed or redundant candidates according to quality dimensions such as clarity,
relevance, and diversity, yielding structured question templates
$\{\mathcal{T}_i\}$ for the next stage.
\DTTightBetween

\DTTightParagraph
\paragraph{Stage \ding{176}: Critic-Guided Question--Answer--Explanation Generation.}
For each template $\mathcal{T}_i$, a \textbf{generator} produces a question, answer, and explanation triple $(q_i, a_i, e_i)$, drawing on $\mathcal{C}_{\text{rag}}$ for factual grounding and $\mathcal{C}_{\text{mem}}$ for difficulty calibration.
A structurally separated \textbf{validator} then applies two complementary checks: LLM-based verification assesses template alignment, factual correctness, and pedagogical soundness for all items, while computational questions undergo additional sandboxed code execution.
This separation is critical: because the validator shares no reasoning chain with the generator, it must independently verify correctness, reducing the risk of self-confirming errors that self-evaluation may miss~\citep{spiliopoulou2025playfavorites}.
Failed pairs receive structured diagnostic feedback and are regenerated until both pedagogical and factual constraints are met.
\DTTightBetween

% ===========================================================================
\subsection{Closed-Loop Tutoring Cycle}
\label{sec:system}
\DTTightSubsection
% ===========================================================================

The components above form a self-reinforcing cycle.
After each interaction, the pipeline appends a new trace tree to $\mathcal{F}$ and the three memory agents update $\mathcal{D}$ in parallel, producing a revised profile that immediately informs subsequent sessions.
The key property is \textit{bidirectional task coupling}, which operationalizes formative assessment: weaknesses diagnosed during problem tutoring propagate to $\mathcal{D}_w$ and directly shape which questions are generated next; conversely, the learner's performance on generated questions refines $\mathcal{D}_s$ and $\mathcal{D}_r$, improving future explanations.
Because every agent can query the trace forest through the TraceToolkit, this feedback extends beyond profile-level summaries to fine-grained precedents from any prior interaction, enabling increasingly nuanced personalization as the history grows. Appendix~\ref{app:prompts} further provides design details of DeepTutor.
\DTTightBetween

\begin{figure*}[!t]
  \DTTightFloatTop
  \centering
  \includegraphics[width=\textwidth]{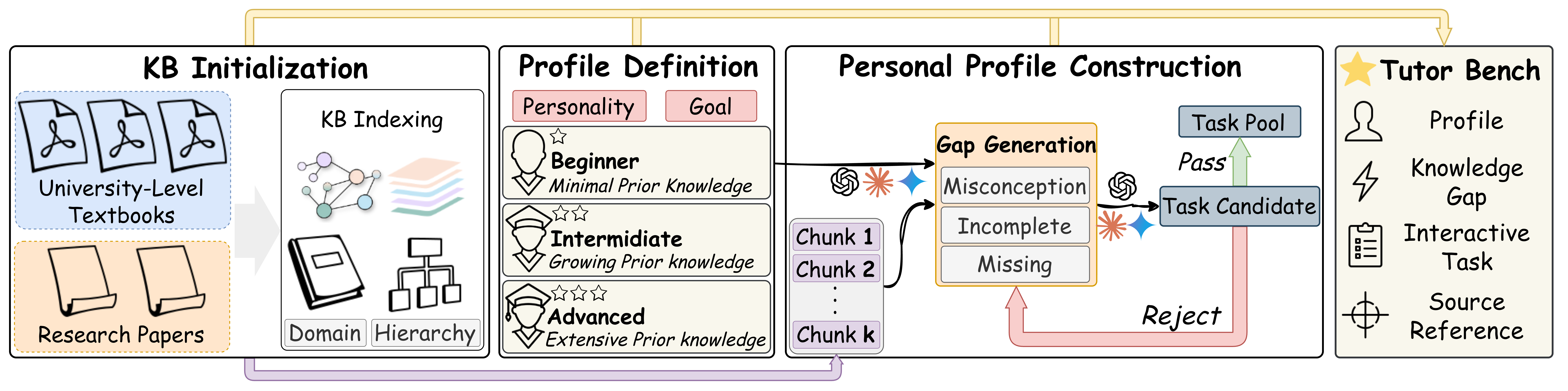}
  \DTTightCaption
  \caption{Construction pipeline of TutorBench. Four stages produce entries each containing a learner profile, grounded knowledge gaps, an interactive task, and source references.}
  \label{fig:datagen}
  \DTTightFloatBottom
\end{figure*}

\section{Extending the Personalization Substrate}
\label{sec:extensions}
\DTTightSection

The previous section focuses on the quantitatively evaluated core of
\textsc{DeepTutor}: citation-grounded problem tutoring, personalized question
generation, and learner-memory updates. In practice, however, learning extends
beyond a sequence of question-answering turns. Students synthesize information
across sources, construct alternative representations of abstract concepts,
externalize understanding through writing, and revisit weak concepts over
longer time horizons~\citep{scardamalia2006knowledge,graham2020writing}.
If each activity is implemented as an independent tool or interface, the
learner model again becomes fragmented. We therefore treat the Hybrid
Personalization Engine as a reusable substrate that can support both new
agentic capabilities and new learner-facing workspaces.
\DTTightBetween

\begin{figure*}[t]
    \DTTightFloatTop
    \centering
    \includegraphics[width=\textwidth]{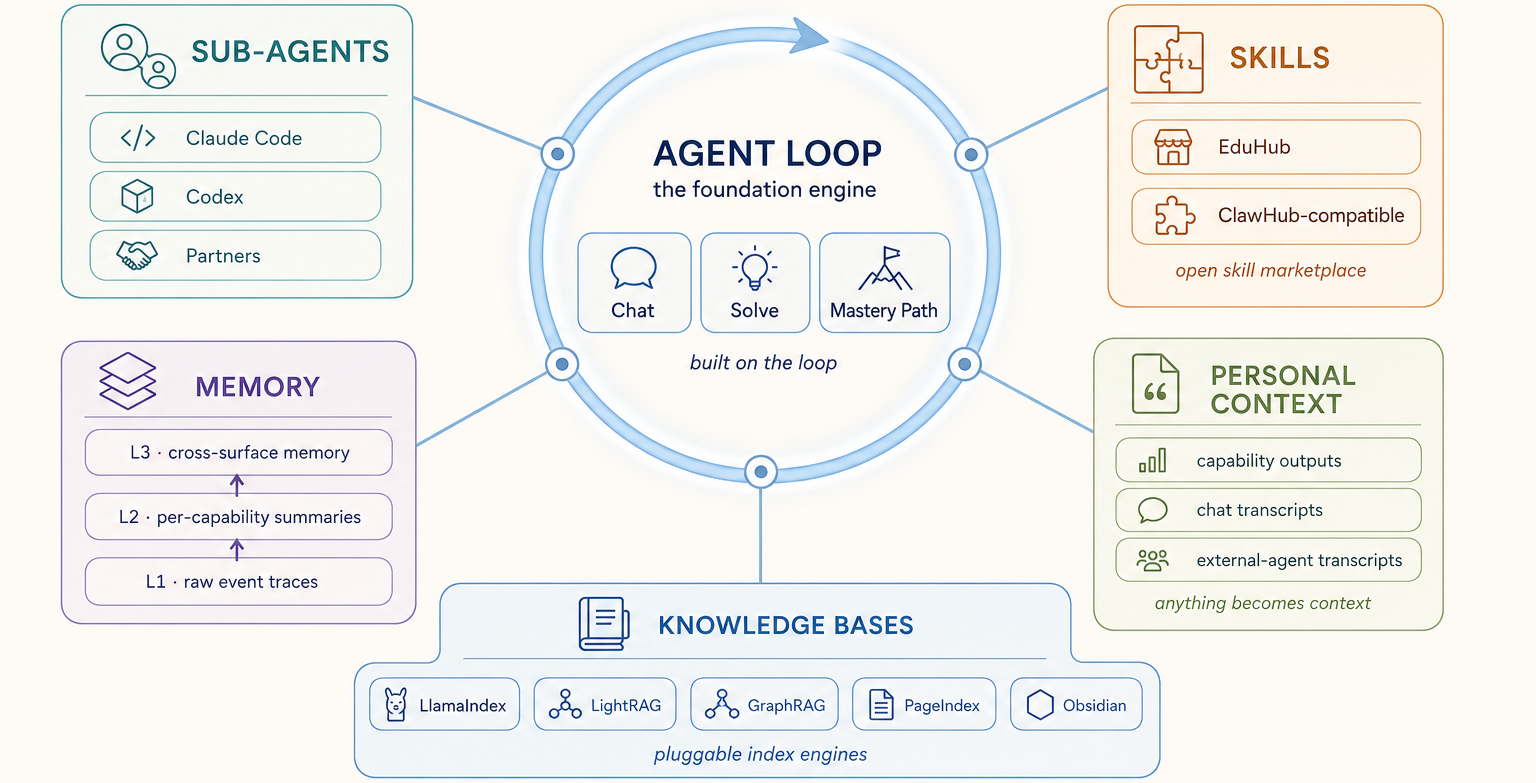}
    \DTTightCaption
    \caption{The Hybrid Personalization Engine as a reusable substrate.
    A single agent loop drives every capability (Chat, Solve, Mastery Path, and the
    extensions of this section), while cross-surface memory (raw event traces,
    per-capability summaries, and cross-surface synthesis over the Trace Forest and
    learner profile), pluggable knowledge-base engines, installable skills,
    consultable sub-agents, and accumulated personal context are shared across all of
    them. Every extension reuses this substrate instead of instantiating a separate
    learner model.}
    \label{fig:substrate}
    \DTTightFloatBottom
\end{figure*}

As illustrated in Figure~\ref{fig:substrate}, this substrate has a small,
fixed set of shared parts. A single agent loop executes every capability, from
the tutoring core to the extensions below; the same cross-surface memory,
built on the Trace Forest and learner profile $\mathcal{D}$ of
\S\ref{sec:memory}, records interactions from all surfaces rather than one;
pluggable knowledge-base engines supply source grounding; installable skills
supply reusable procedures; consultable sub-agents supply external expertise;
and accumulated personal context feeds back into every subsequent turn.
Each extension in this section is an instantiation over these shared parts, not
a parallel system with its own learner model.
Table~\ref{tab:extensions} summarizes how the main system extensions reuse this
substrate at three levels: capability execution, learning surfaces, and
long-horizon deployment.
\DTTightBetween

\begin{table*}[!t]
\DTTightFloatTop
\centering
\small
\setlength{\tabcolsep}{4pt}
\renewcommand{\arraystretch}{1.04}
\begin{tabularx}{\textwidth}{>{\raggedright\arraybackslash}p{0.14\textwidth}
                              >{\raggedright\arraybackslash}p{0.17\textwidth}
                              >{\raggedright\arraybackslash}X
                              >{\raggedright\arraybackslash}X}
\toprule
\textbf{Level} & \textbf{Instantiation} & \textbf{Relation to Core Tutoring} & \textbf{Reused Substrate} \\
\midrule
Capability runtime & Deep Research &
First-class capability for open-ended knowledge synthesis, parallel to Deep Solve. &
Query decomposition; KB/web/paper search; tool execution; report synthesis. \\
Capability runtime & Visualize / Math Animator &
First-class representation capability for diagrams, charts, animations, and demos. &
Dialogue context; visual-code generation and review; rendering loop. \\
Capability runtime & Subagent (My Agents) &
Consults a live external agent, or imports its transcripts, without reimplementing its expertise. &
Turn context; \texttt{consult\_subagent} tool (bounded rounds); imported transcripts as context. \\
Learning surface & Mastery Path &
Long-horizon practice plan that sequences and reschedules review across sessions. &
Learner profile and weak concepts; scheduling; grading; question generation. \\
Learning surface & Co-Writer &
Document-editing page with embedded AI writing actions, not another chat capability. &
Markdown workspace; selected spans; optional KB/web grounding; rewrite/expand/shorten with accept/reject diffs. \\
Curriculum artifact & Book Engine &
Composes sources, prior interactions, and capability outputs into an interactive book. &
Input snapshot; source exploration; spine and concept graph; typed blocks; per-page chat; quiz progress. \\
Multi-channel deployment & Partners &
Extends the capability layer beyond the web into proactive, channel-based agents (TutorBot runtime). &
Persona, skills, heartbeat scheduling, channel adapters, shared capabilities. \\
\bottomrule
\end{tabularx}
\DTTightTableCaption
\caption{System extensions supported by the same personalization substrate.}
\label{tab:extensions}
\DTTightFloatBottom
\end{table*}

% ===========================================================================
\subsection{From Tutoring Workflows to Adaptive Learning}
\label{sec:extensions:adaptive-learning}
\DTTightSubsection
% ===========================================================================

The experimentally evaluated core of \textsc{DeepTutor} personalizes two
tutoring capabilities: citation-grounded problem tutoring and
difficulty-calibrated question generation. In the implemented system, these
capabilities are part of a broader capability runtime rather than isolated
features. Deep Research, Visualize, and subagent consultation extend the same runtime
laterally: they are invoked as first-class capabilities alongside Deep Solve and
Question Generation, but target different learning needs. Mastery Path and
Co-Writer extend the system in a different way: the former turns single-turn
tutoring into a rescheduled long-horizon practice plan, while the latter is a
Markdown editing workspace with embedded AI actions, so the learner can work on
persistent documents while selectively invoking context-aware assistance. This
distinction is important because adaptive learning requires both callable
agentic capabilities and durable surfaces where learners externalize, practice,
and refine their understanding.
\DTTightBetween

\DTTightParagraph
\paragraph{Adaptive knowledge synthesis.}
Open-ended learning tasks often require more than answering a single prompt:
the learner must identify subtopics, gather evidence, compare sources, and
integrate findings into a coherent view~\citep{bereiter2002education}. The
Deep Research capability operationalizes this process through a multi-stage
agentic pipeline. It first reformulates the learner's request, decomposes it
into research directions, dispatches research agents over course materials, web
sources, academic-paper search, and computational tools, and finally synthesizes
the evidence into a structured report. Because it shares the same runtime as
Deep Solve and Question Generation, the resulting research trajectory can be
interpreted as evidence about the learner's interests, recurring confusions,
and emerging knowledge goals.
\DTTightBetween

\DTTightParagraph
\paragraph{Adaptive access to external expertise.}
Some learning tasks are best served by expertise that already lives outside the
tutor, such as a coding agent running on the learner's own machine. The subagent
capability, surfaced to learners as \emph{My Agents}, lets a chat turn consult a
live external agent, for example a Claude Code or Codex session, or a Partner
(\S\ref{sec:extensions:tutorbot}), and stream its work back into the same turn;
past transcripts of such agents can also be imported as first-person references.
Consultation is selected like any other grounding source: the tutor runs a
bounded number of consults, treats the returned work as evidence, and then
answers in its own voice. External expertise therefore enters the learner model
as additional personal context, rather than as a separate ungrounded channel.
\DTTightBetween

\DTTightParagraph
\paragraph{Adaptive knowledge representation.}
Textual explanation alone is often insufficient when the target concept is
structural, procedural, or geometric. The Visualize and Math Animator
capabilities therefore treat representation choice as a pedagogical action
rather than a surface formatting step. They analyze the conceptual target and
surrounding dialogue, generate executable visual artifacts such as diagrams,
charts, animations, or interactive demonstrations, and review the artifact
before returning it to the learner. In this view, visualization is selected
when the learner state or the content structure suggests that an alternative
representation may reduce cognitive load or expose relations hidden in
prose~\citep{ainsworth2006deft,mayer2003promise}.
\DTTightBetween

\DTTightParagraph
\paragraph{Adaptive knowledge expression.}
Writing provides a complementary diagnostic surface: students reveal conceptual
gaps not only by giving wrong answers, but also by omitting prerequisites,
using unstable terminology, or failing to connect claims causally. Co-Writer is
therefore implemented as a Markdown editor rather than as another chat
workflow. The learner works directly with documents and selected spans, while
AI actions can rewrite, expand, shorten, or reorganize text with optional
course-material or web grounding. Because each action is applied to a selected
span and returned as an accept/reject diff, the learner controls what lands in
the document, and finished drafts can be saved back into notebooks as reusable
context. This interface turns writing into a durable learning surface: document
revisions and selected passages can become evidence for later tutoring, while
assistance remains grounded in the same learner and source context.
\DTTightBetween

\DTTightParagraph
\paragraph{Adaptive practice over long horizons.}
Diagnosis and explanation address the current turn, but durable learning also
requires spaced, sequenced practice. Mastery Path turns a target topic into a
learning plan whose steps are sequenced, graded, and rescheduled across
sessions: each attempt is graded against a reference, updates the weak-concept
inventory $\mathcal{D}_w$, and feeds a scheduler that decides what to revisit and
when. Because the plan draws practice items from the same question-generation
capability and reads the same learner profile, mastery progress and tutoring
history stay coupled rather than tracked in a separate mastery model.
\DTTightBetween

Together, these extensions move personalization from local tutoring turns to a
broader adaptive learning process. Deep Research, Visualize, and subagent
consultation expand the capability runtime; Mastery Path and Co-Writer add
durable practice and authoring surfaces. The common point is that synthesis,
representation, practice, and expression can share the same learner-aware
substrate instead of introducing separate learner models for each activity.
\DTTightBetween

% ===========================================================================
\subsection{Interactive Books through Substrate Reuse}
\label{sec:extensions:books}
\DTTightSubsection
% ===========================================================================

The Book Engine extends the same principle from individual workflows to a
curricular artifact. Whereas a tutoring turn produces an answer and a practice
request, an interactive book compiles a learner-conditioned collection of
sources, prior interactions, notes, and question entries into a persistent
learning environment. This design is motivated by the observation that
long-horizon learning requires not only local feedback, but also organized
paths through concepts, prerequisites, examples, and self-checks.
Recent work on AI-augmented textbooks similarly highlights the value of
source-faithful personalization, multiple representations, and embedded
assessment in transforming static textbooks into adaptive learning
experiences~\citep{team2025towards}.
\DTTightBetween

\DTTightParagraph
\paragraph{Input snapshot and proposal.}
The book workflow begins by freezing a multi-source learning context rather
than generating from a bare title. The current implementation accepts user
intent, selected chat history, notebook references, knowledge bases, question
categories, and question entries as book inputs. An ideation stage turns this
input snapshot into a proposal specifying the book's scope, audience,
pedagogical objective, and expected structure. This explicit proposal stage
keeps curriculum construction inspectable before expensive generation begins.
\DTTightBetween

\DTTightParagraph
\paragraph{Curricular spine and concept graph.}
After proposal confirmation, a source-exploration stage performs a targeted
sweep over the selected materials. A spine synthesizer then constructs the
chapter structure and concept graph, making prerequisite relations and topical
coverage explicit before page-level compilation. This step reuses the same
source-grounding principle as the tutoring core, but changes the unit of
organization from a dialogue turn to a curriculum path.
\DTTightBetween

\DTTightParagraph
\paragraph{Block-level page compilation.}
Each page is generated as an ordered sequence of typed blocks rather than a
monolithic document. Textual explanation, callouts, quizzes, figures,
interactive demonstrations, animations, code, timelines, flash cards, deep-dive
links, section summaries, and concept-graph blocks can be selected according to
the pedagogical role of the content. This representation makes the book
compatible with the adaptive capabilities and surfaces above: research supplies
evidence, visualization supplies representational blocks, question generation
supplies self-checks, and the writing surface supports learner-authored notes.
\DTTightBetween

\DTTightParagraph
\paragraph{Persistent interaction and remediation.}
The book remains editable and interactive after compilation. Individual blocks
can be regenerated or inserted, deep-dive subpages can be spawned from a concept,
each page carries its own page-level chat, and quiz attempts update progress
records such as visited pages, bookmarks, scores, and weak chapters.
Consequently, the book acts less like a static export and more like a persistent
learning workspace. Because compiled pages are fingerprinted against their
sources, the system can also detect when source knowledge has drifted from the
book and flag the affected pages for refresh. Its interaction logs and
weak-chapter signals can provide additional evidence for future learner-profile
updates, while the profile can guide subsequent supplementation or remediation.
\DTTightBetween

The Book Engine therefore illustrates a stronger form of substrate reuse:
\textsc{DeepTutor} does not merely call tutoring capabilities inside a document
viewer; it composes source exploration, concept organization, multimodal block
generation, question-based self-checking, and progress tracking into a single
adaptive learning artifact.
\DTTightBetween

% ===========================================================================
\subsection{Proactive Multi-channel Tutoring}
\label{sec:extensions:tutorbot}
\DTTightSubsection
% ===========================================================================

The previous extensions remain primarily learner-initiated: the student chooses
when to request research, visualization, writing support, practice, or a book.
Long-term learning, however, also depends on timely review and low-friction
follow-up. Reactive interfaces require the learner to remember when to return,
choose an entry point, and formulate the next request. \textsc{Partners} address
this deployment problem by reusing the same tutoring capabilities inside
persistent, multi-channel agent instances. A Partner is deliberately not a
separate bot engine: every inbound web or messaging event becomes an ordinary
orchestrator turn inside a Partner-scoped workspace, driven by the same
autonomous agent loop (the \textsc{TutorBot} runtime detailed in
Appendix~\ref{app:tutorbot}). Informally, a Partner is a chat that also carries a
persona and a channel address.
\DTTightBetween

\DTTightParagraph
\paragraph{Persistent agent instances.}
Each Partner maintains its own workspace, conversation state, persona
specification, skill set, and long-term memory while sharing the underlying
\textsc{DeepTutor} capability layer. This separation supports specialized
agents, such as a mathematics tutor, research assistant, or exam-preparation
coach, without duplicating the tutoring runtime. A Partner reads its owner's
memory but writes only its own, and a message adapter converts incoming channel
events into the unified context consumed by the orchestrator, so
Partner-initiated interactions and ordinary tutoring sessions remain compatible.
\DTTightBetween

\DTTightParagraph
\paragraph{Capability reuse through skills.}
Partners do not reimplement tutoring behavior. Their skills describe when and
how to invoke high-level actions such as grounded explanation, problem solving,
question generation, research assistance, visualization, document interaction,
or scheduling. Skills therefore act as workflow descriptors over the same
tooling and personalization substrate, allowing proactive agents to compose
validated capabilities rather than rely on a separate prompt-only interface.
Skills use an open, self-contained format and can be shared through a community
registry (EduHub), so the same substrate can be extended with new procedures
without modifying the core system, and a skill authored for one surface remains
usable across chat, books, and Partners alike.
\DTTightBetween

\DTTightParagraph
\paragraph{Heartbeat, scheduling, and channels.}
A heartbeat and scheduling layer allows a Partner to periodically decide whether
proactive action is pedagogically useful, for example by reminding the learner
to revisit an active weakness, generating a short review task after a delay, or
summarizing newly added material. Schema-driven channel adapters route
interactions from more than a dozen external messaging platforms, including
Feishu, Slack, Discord, Telegram, WhatsApp, and Microsoft Teams, into the same
backend context. The goal is not to multiply interfaces for their own sake, but
to reduce context fragmentation: a learner can interact across devices while the
tutor preserves a coherent history and learner model.
\DTTightBetween

We position Partners as a deployment mechanism for long-horizon personalized
tutoring rather than as an additional intervention validated by the present
benchmark. Their effect on retention, engagement, interruption cost, and real
learning outcomes requires longitudinal human studies.
\DTTightBetween

% ===========================================================================
% Section 4: TutorBench
% ===========================================================================

\section{TutorBench}
\label{sec:tutorbench}
\DTTightSection

Existing educational benchmarks predominantly adopt an \emph{instructor-centric} perspective, while treating the student as a generic receiver~\citep{chu-etal-2025-llm}.
The few learner-aware efforts model students through coarse labels (e.g., ``struggling'' vs.\ ``proficient'') without grounding profiles in the actual course content~\citep{park2024empoweringpersonalizedlearning,wang2025llmpoweredmultiagentframeworkgoaloriented}.
Therefore, we construct TutorBench to close this gap: every entry explicitly couples a detailed learner persona with source-grounded knowledge gaps and an interactive tutoring task, as illustrated in Figure~\ref{fig:datagen}.
\DTTightBetween

\DTTightParagraph
\paragraph{Construction Pipeline.}
University-level textbooks and research papers are first indexed into the dual-representation knowledge base (\S\ref{sec:rag}), from which a domain hierarchy is derived.
For each KB, we instantiate three learner profiles using \textit{beginner}, \textit{intermediate}, and \textit{advanced} only as high-level starting points; each profile is further grounded in educational background, learning purpose, and fine-grained per-topic mastery states: \textit{known well}, \textit{partially known}, and \textit{unknown}.
For each profile, we sample source pages and generate three types of grounded knowledge gaps: \textit{misconceptions}, \textit{incomplete understanding}, and \textit{missing knowledge}~\citep{smith1993misconceptions,shute2008focus}.
Each gap is anchored to specific source pages and includes both a manifestation (how the student would \textit{exhibit} it) and the correct understanding serving as evaluation reference.
During evaluation, all tutor systems receive the same task interface and access to the same source KB. The learner profile and gap manifestations are used to initialize the student simulator, while the correct understandings are reserved for judging and are not exposed to tutor systems as ground-truth answers.
Interactive tasks are generated around selected gaps via rejection sampling, which verifies gap coherence, task--gap alignment, and conversational naturalness; the full procedure is provided in Appendix Algorithm~\ref{alg:task}.
All generated profiles and tasks then undergo a final human review, where annotators verify factual correctness, profile plausibility, gap--task consistency, and pedagogical naturalness before inclusion in the benchmark.
Full construction details are provided in Appendix~\ref{app:tutorbench}.
\DTTightBetween

\DTTightParagraph
\paragraph{Statistics.}
TutorBench draws on 30 knowledge bases spanning five broad disciplines, covering \textit{humanities, sciences, engineering, business, and frontier research}.
Based on this coverage, we instantiate three student profiles with different levels for each knowledge base, yielding 90 student profiles grounded in explicit KB materials.
After rejection sampling, we retain exactly three accepted interactive tasks for each profile, resulting in 270 tasks in total. All retained entries are further checked by human reviewers before final release.
These tasks cover concept understanding, problem solving, application, and comparison-oriented tasks.
Every entry carries three source-grounded knowledge gaps, each anchored to specific pages from the source KB materials.

% ===========================================================================
% Section 4: Experiments
% ===========================================================================

\section{Experiments}
\label{sec:experiments}
\DTTightSection

We evaluate DeepTutor along five complementary axes:
\emph{first-person interactive evaluation} (\S\ref{sec:interactive_eval}),
\emph{cross-domain generalization} (\S\ref{sec:cross_domain}),
\emph{human preference alignment} (\S\ref{sec:human_alignment}),
\emph{component-level ablation} (\S\ref{sec:ablation}),
and \emph{extended generalization of agentic problem solving} (\S\ref{sec:generalization}).
The system extensions in Section~\ref{sec:extensions} demonstrate how the same
substrate supports broader adaptive learning workflows, interactive books, and
proactive agents. Our quantitative evaluation focuses on the foundational
closed-loop tutoring core, where controlled benchmarking is feasible;
longitudinal evaluation of the interactive-book and Partners layers is left for
future work.
\DTTightBetween

{%
\setlength{\aboverulesep}{0pt}%
\setlength{\belowrulesep}{0pt}%
\renewcommand{\arraystretch}{1.18}%
\begin{table*}[!t]
\DTTightFloatTop
\centering
\small
\setlength{\tabcolsep}{4.2pt}
\resizebox{\textwidth}{!}{%
\begin{tabular}{l ccccc g ccccc g cc}
\toprule
& \multicolumn{6}{c}{\textbf{Tutoring Quality}}
& \multicolumn{6}{c}{\textbf{Practice Quality}}
& \multicolumn{2}{c}{\textbf{Overall}} \\
\cmidrule(lr){2-7} \cmidrule(lr){8-13} \cmidrule(lr){14-15}
\textbf{System}
  & SF & PER & APP & VID & LD & \textit{Avg}
  & FIT & GND & DIV & ANS & CC & \textit{Avg}
  & OQ & $\Delta$\% \\
\midrule
Naive Tutor
  & 3.22 & 4.24 & 4.44 & 3.89 & 3.99 & 3.96
  & \underline{3.20} & 2.46 & 2.83 & 3.87 & \underline{3.12} & \underline{3.10}
  & 3.53 & \textcolor{gray}{--} \\
CoT Tutor
  & 3.34 & 4.22 & 4.47 & 3.82 & 4.01 & 3.97
  & 3.18 & \underline{2.50} & 2.79 & 3.81 & 3.04 & 3.06
  & 3.52 & {-0.28\%} \\
Self-Refine Tutor
  & 3.28 & 4.28 & \textbf{4.61} & \underline{3.94} & \underline{4.14} & \underline{4.05}
  & 3.17 & 2.49 & \underline{2.88} & 3.88 & 2.97 & 3.08
  & \underline{3.57} & \underline{+1.13\%} \\
ReAct Tutor
  & \underline{3.35} & \underline{4.37} & 4.40 & 3.70 & 3.96 & 3.96
  & 3.16 & 2.47 & 2.75 & \underline{3.94} & 3.07 & 3.08
  & 3.52 & {-0.28\%} \\
\midrule
\textsc{DeepTutor}
  & \textbf{3.36} & \textbf{4.59} & \underline{4.56} & \textbf{4.81} & \textbf{4.61} & \textbf{4.39}
  & \textbf{3.35} & \textbf{2.96} & \textbf{3.44} & \textbf{3.98} & \textbf{3.38} & \textbf{3.42}
  & \textbf{3.91} & \textbf{+10.76\%} \\
\bottomrule
\end{tabular}
}
\DTTightTableCaption
\caption{Main results of interactive evaluation. \textit{Avg} is the group mean, OQ means \underline{O}verall \underline{Q}uality scores across ten metrics. $\Delta$\% is relative improvement over Naive Tutor. All metrics range from 1 to 5 $\uparrow$. \textbf{Bold} represents the best scores, \underline{underline} represents the second.}
\label{tab:interactive_all}
\DTTightFloatBottom
\end{table*}
}

% ===========================================================================
\subsection{First-Person Interactive Evaluation}
\label{sec:interactive_eval}
\DTTightSubsection
% ===========================================================================
\begin{wrapfigure}{r}{0.48\textwidth}
  \centering
  \includegraphics[width=\linewidth]{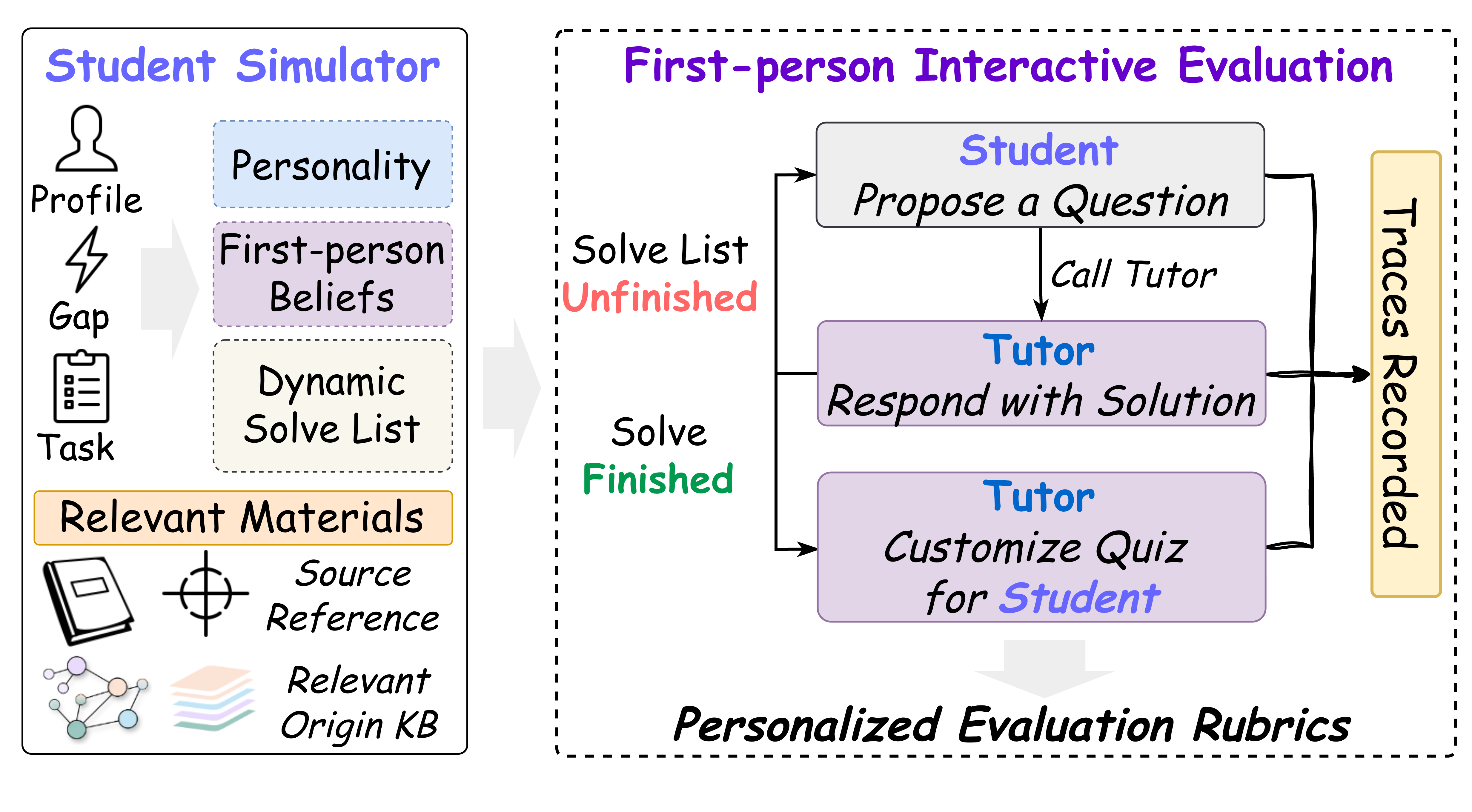}
  \vspace{-10pt}
  \caption{TutorBench first-person interactive evaluation protocol.}
  \label{fig:eval_protocol}
  \vspace{-10pt}
\end{wrapfigure}
The adaptive diagnosis and guided remediation that unfold over multiple turns are often missed by traditional single-turn metrics.
Building on rubric-based LLM evaluation in open-ended settings~\citep{zheng2023judging,huang2025mathtutorbench} and simulated-student paradigms for tutoring assessment and training~\citep{dinucujianu2025pedagogy,scarlatos2026simulatedstudents,pan2025tutorup}, we introduce a TutorBench protocol in which an LLM-based student simulator engages each tutoring system, and the resulting transcripts are scored against personalized rubrics.
Figure~\ref{fig:eval_protocol} summarizes the loop: a benchmark entry initializes the learner profile, source-grounded gaps, task, and materials; the simulator converts the gaps into first-person beliefs, conducts multi-turn tutoring, and then requests customized practice; the resulting traces are judged with personalized rubrics.
This design lets the simulator embody the student's persona rather than merely narrating errors from an external perspective~\cite{scarlatos2026simulatedstudents}, while dynamic gap pacing elicits resistance patterns that reflect sustained tutoring interactions~\cite{pan2025tutorup}.
\DTTightBetween

\DTTightParagraph
\paragraph{Metrics.}
To evaluate tutoring from both learner-facing explanation and follow-up practice perspectives, we draw on established work in formative feedback, learner modeling, source-grounded generation, and educational assessment~\citep{black1998assessment,shute2008focus,corbett1994knowledge,anderson1995cognitive,kurdi2020systematic}.
This leads to ten complementary dimensions, each scored 1--5 and organized into two groups, rather than a single aggregate quality label.
Five \textbf{tutoring-side} metrics evaluate the response quality every turn.
\textit{Source Faithfulness}~(SF) captures factual consistency with the source material and attribution to provided evidence~\citep{maynez2020faithfulness,rashkin2023measuring}.
\textit{Personalization}~(PER) reflects how specifically the response addresses the learner's current confusion while remaining aligned with the learner state developed over the session~\citep{vygotsky1978mind,corbett1994knowledge}.
\textit{Applicability}~(APP) evaluates whether the guidance is concrete and immediately actionable~\citep{hattie2007power,shute2008focus}.
\textit{Vividness}~(VID) rewards rich, engaging presentation through effective use of multiple representations beyond plain prose~\citep{ainsworth2006deft,mayer2003promise}.
\textit{Logical Depth}~(LD) assesses the presence of explicit causal chains and intermediate reasoning over bare assertions~\citep{chi1989self,koedinger2012knowledge}.
Five \textbf{practice-side} metrics score the generated questions after each dialogue.
\textit{Fitness}~(FIT) measures alignment with session-level diagnosed weaknesses at appropriate difficulty~\citep{wood1976role,lord1980applications}.
\textit{Groundedness}~(GND) requires the question stem, answer, and explanation to be factually anchored in the source material~\citep{lewis2020retrieval,gao2023enabling}.
\textit{Diversity}~(DIV) rewards novelty in angle and cognitive demand~\citep{anderson2001taxonomy,kurdi2020systematic}.
\textit{Answer Quality}~(ANS) evaluates correctness of the key and plausibility of distractors~\citep{haladyna2002review}.
\textit{Cross Concept}~(CC) tests whether the question meaningfully integrates multiple concepts surfaced throughout the session~\citep{chi1981categorization,barnett2002when}.
\DTTightBetween

\DTTightParagraph
\paragraph{Protocol.}
Operationally, the evaluation follows three stages: offline entry generation, multi-turn simulation, and LLM-based independent judging.
To mitigate the instability of LLM-based evaluation, each transcript is scored three times by a fixed judge and the scores are averaged. For the reported interactive evaluation, we use \textit{Gemini-3-Flash} to power the student simulator as well as all tutor backbones including DeepTutor, while \textit{Claude Sonnet~4.6} serves as the judge with temperature set to zero.
All reported interactive results are obtained on the full TutorBench release, covering 270 tasks instantiated from 90 student profiles grounded in 30 knowledge bases.
For reproducibility, Appendix~\ref{app:eval:interactive} gives the full first-person evaluation details, including entry construction, simulator and judge prompts, scoring protocol, and exact rubric wording.
\DTTightBetween
\DTTightParagraph
\paragraph{Baselines.}
Existing LLM-agent systems for education and open-domain reasoning vary substantially in scope, and several tutoring systems are either unavailable as open-source releases or difficult to reproduce end-to-end in our setting.
We therefore design four representative tutor baselines within one unified harness. They share the same RAG tool and backbone with \textsc{DeepTutor}. Among them, \textit{Naive Tutor} uses direct prompting; \textit{CoT Tutor} adds chain-of-thought reasoning~\citep{wei2022chain}; \textit{Self-Refine Tutor} adds a pedagogical review pass~\citep{madaan2024selfrefine}; and \textit{ReAct Tutor} implements a \textit{thought-act-observe} reasoning loop~\citep{yao2023reactsynergizingreasoningacting}.
\DTTightBetween

\DTTightParagraph
\paragraph{Results.}
Main experimental results are reported in Table~\ref{tab:interactive_all}.
The four baselines remain tightly clustered, indicating that adding CoT, self-refinement, or ReAct-style tool use to the same backbone and RAG interface is not sufficient to match \textsc{DeepTutor}'s learner-adaptive behavior.
\textsc{DeepTutor} improves overall quality by 10.76\% and leads on nearly all metrics.
On the tutoring side, the most visible gains appear in \textit{Vividness}, \textit{Personalization}, and \textit{Logical Depth}, matching the method's separation of investigation, guided solving, and adaptive writing: the system first diagnoses the gap, then builds a causal solution path, and finally presents it at an appropriate level of detail.
\textit{Source Faithfulness} and \textit{Applicability} remain competitive, suggesting that personalization does not come at the cost of grounding or actionability.
On the practice side, large gains in \textit{Groundedness}, \textit{Diversity}, and \textit{Cross Concept} indicate stronger source anchoring and concept coverage, while smaller gains in \textit{Fitness} and \textit{Answer Quality} suggest more targeted and reliably checked items.
\DTTightBetween

% ===========================================================================
\subsection{Cross-Domain Generalization}
\label{sec:cross_domain}
\DTTightSubsection
% ===========================================================================

{%
\begin{figure*}[!t]
\DTTightFloatTop
\centering
\includegraphics[width=\textwidth]{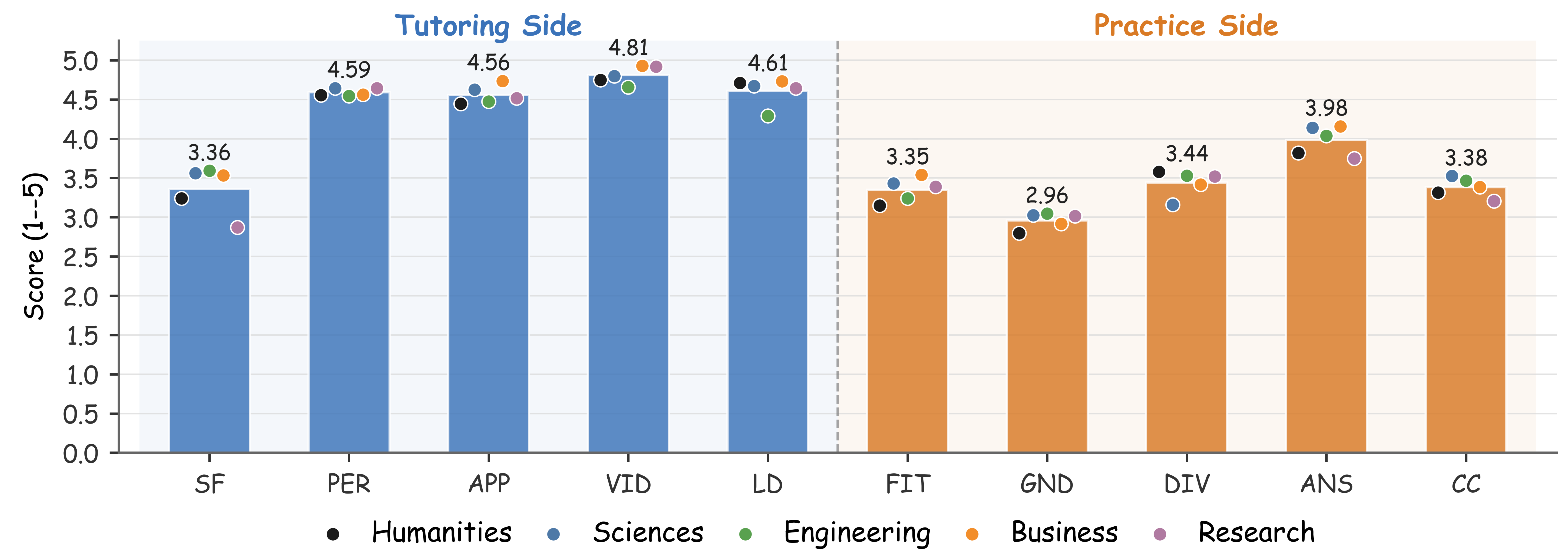}
\DTTightCaption
\caption{Cross-domain metric profiles for the first-person interactive simulation. Bars show the average score across the five domains, while colored dots show domain-specific scores.}
\label{fig:domain}
\DTTightFloatBottom
\end{figure*}
}

We further decompose \textsc{DeepTutor}'s first-person interactive performance by metric and discipline in Figure~\ref{fig:domain}, with the complete domain-by-metric table in Appendix~\ref{app:results:breakdown}.
This analysis tests whether the investigate--solve--write scaffold depends on one content area, or remains stable across materials with different discourse structures and evidence density.
Overall quality varies by only 0.16 points across the five domains, indicating that the gains are not driven by a single discipline.
At the metric level, the average cross-domain span is 0.36 points, with the largest fluctuation appearing in \textit{Source Faithfulness}; this suggests that evidence anchoring is more domain-sensitive than learner adaptation.
On the tutoring side, \textit{Personalization}, \textit{Applicability}, \textit{Vividness}, and \textit{Logical Depth} are consistently strong, while \textit{Source Faithfulness} is the main relative bottleneck.
On the practice side, \textit{Answer Quality} is strongest, whereas \textit{Groundedness} remains the lowest-scoring dimension, suggesting that generating correct practice items is easier than making every question and explanation explicitly traceable to source evidence.
\DTTightBetween

% ===========================================================================
\subsection{Human Evaluation of Preference Alignment}
\label{sec:human_alignment}
\begin{figure}[!t]
  \DTTightFloatTop
  \centering
  \includegraphics[width=\textwidth]{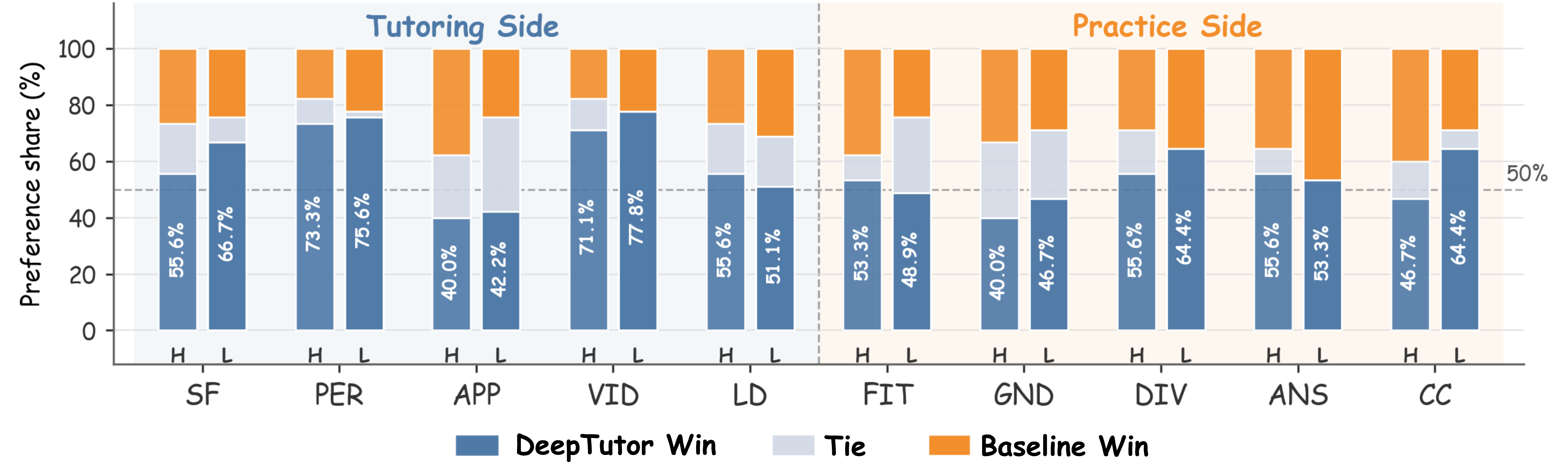}
  \DTTightCaption
  \caption{Human and LLM preference outcomes across ten metrics. H/L denote human-majority and LLM-judge preferences; colors indicate \textsc{DeepTutor} win, tie, and baseline win.}
  \label{fig:human_alignment}
  \DTTightFloatBottom
\end{figure}

\DTTightSubsection
% ===========================================================================
\begin{wrapfigure}{r}{0.48\textwidth}
  \centering
  \includegraphics[width=\linewidth]{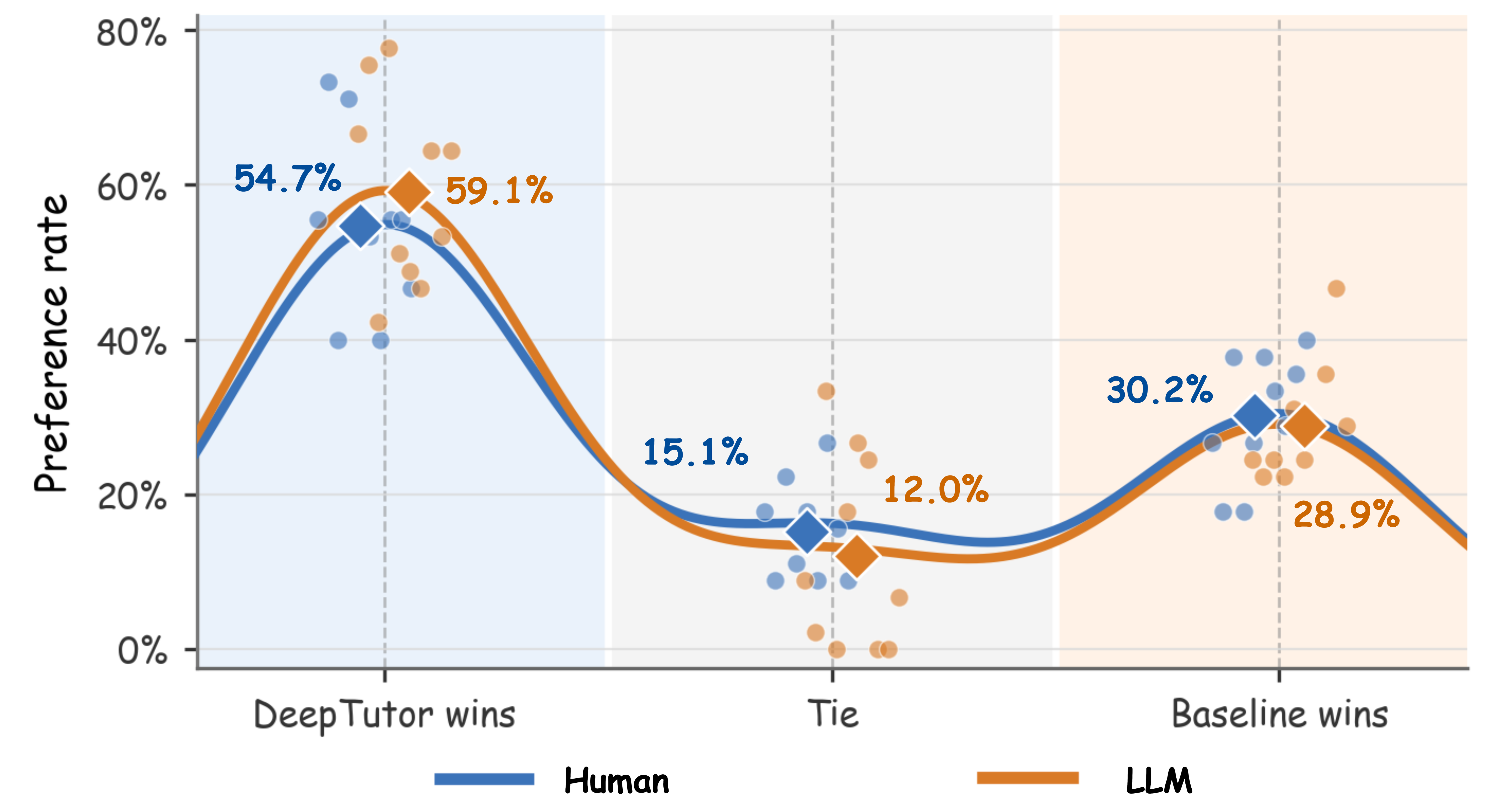}
  \vspace{-12pt}
  \caption{Distribution-level W/T/L alignment between human raters and the LLM judge.}
  \label{fig:preference_align_distribution}
  \vspace{-10pt}
\end{wrapfigure}
To validate whether automated preferences reflect human judgments, we conduct a blind pairwise study on a domain-stratified subset of 45 TutorBench sessions, with nine sessions sampled from each domain.
For each selected benchmark entry, human raters compare anonymized \textsc{DeepTutor} and \textit{Naive Tutor (Mock)} outputs generated under the same learner profile, task, source material, student-simulator protocol, and practice-generation request.
The comparison packet includes the learner profile, task description, source material, tutoring transcript, and generated practice questions, so raters can judge both tutoring behavior and follow-up practice quality.
Raters choose the better system on each metric or mark a tie; disagreements between the two assigned annotators for a session--metric pair are conservatively aggregated as ties.
Appendix~\ref{app:eval:human_alignment} documents the annotator instructions, aggregation rules, and additional agreement analyses.
\DTTightBetween
Figures~\ref{fig:preference_align_distribution} and~\ref{fig:human_alignment} compare automated and human preferences at distribution and metric levels.
Both judges assign the largest preference share to \textsc{DeepTutor} across all ten metrics, with the clearest margins on Personalization, Vividness, and Diversity.
More contested metrics such as Applicability, Groundedness, and Cross Concept yield more ties or narrower margins, but the trends remain aligned: human and LLM \textsc{DeepTutor} win rates are strongly correlated across the ten metric-level win-rate pairs (Pearson $r=0.82$, $p=0.0038$; Spearman $\rho=0.83$, $p=0.0027$).
This agreement suggests that the LLM judge is not merely favoring \textsc{DeepTutor} globally, but ranks the systems' relative strengths and weaknesses similarly to human raters under the proposed rubric.
\DTTightBetween

\begin{figure}[!t]
  \DTTightFloatTop
  \centering
  \includegraphics[width=\textwidth]{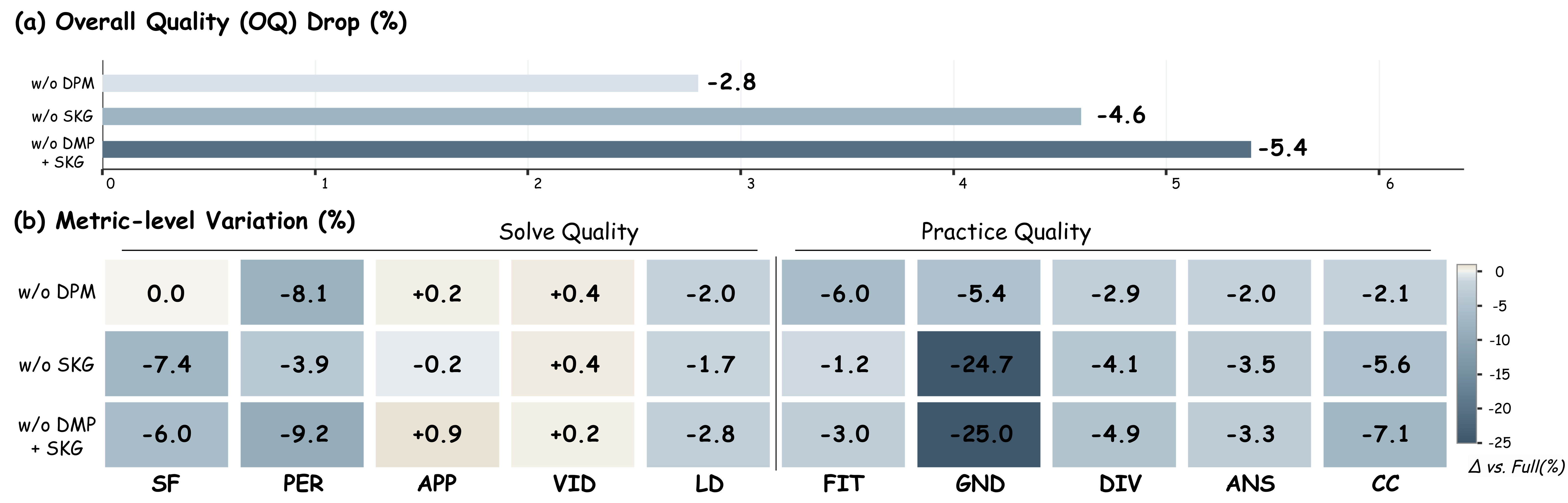}
  \DTTightCaption
  \caption{Component ablation of \textsc{DeepTutor}. Panel (a) shows the overall quality (OQ) drop from removing each module, while panel (b) breaks down the per-metric change relative to the full system.}
  \label{fig:ablation_radar}
  \DTTightFloatBottom
\end{figure}

\DTTightBetween

% ===========================================================================
\subsection{Ablation Study}
\label{sec:ablation}
\DTTightSubsection
% ===========================================================================

We ablate Static Knowledge Grounding (SKG), Dynamic Personal Memory (DPM), or both modules from the full pipeline to isolate their contributions to personalized tutoring.
Figure~\ref{fig:ablation_radar} summarizes both the overall quality degradation and the per-metric changes relative to the full \textsc{DeepTutor} system.
\DTTightBetween

\textbf{Removing SKG} mainly weakens grounding.
\textit{Groundedness} drops the most, followed by \textit{Source Faithfulness} and \textit{Cross Concept}, because both explanations and generated questions lose direct access to source evidence for factual anchoring and cross-reference.
\textbf{Removing DPM} mainly weakens adaptation.
\textit{Personalization} and \textit{Fitness} decline most sharply, showing that the learner profile is central to gap-aware explanation and difficulty-calibrated practice; the smaller drop in \textit{Groundedness} indicates that learner context also helps steer retrieval toward relevant evidence.
The two ablations therefore affect different parts of the rubric: SKG anchors \emph{what} the tutor says, while DPM shapes \emph{how} it adapts. Removing both modules yields the largest overall degradation, supporting SKG and DPM as complementary mechanisms rather than interchangeable add-ons.

% ===========================================================================
\subsection{Extended Generalization of Agentic Problem Solving}
\label{sec:generalization}
\DTTightSubsection
% ===========================================================================

Finally, we test whether DeepTutor's investigate--solve--write scaffold is only useful for personalized tutoring or also transfers to general agentic problem solving.
To isolate this effect, we disable the Hybrid Personalization Engine, including SKG and DPM, and evaluate the remaining solver-only pipeline on five public benchmarks: HLE~\citep{phan2025hle}, GPQA-Diamond~\citep{rein2024gpqa}, LiveBench~\citep{white2025livebench}, GAIA~\citep{mialon2024gaia}, and AA-LCR~\citep{artificialanalysis2025lcr}.
\DTTightBetween

\begin{table*}[!t]
\DTTightFloatTop
\centering
\small
\setlength{\tabcolsep}{5pt}
\resizebox{\textwidth}{!}{%
\begin{tabular}{l ccc cccc c c}
\toprule
& \multicolumn{3}{c}{\textbf{STEM Reasoning}} & \multicolumn{4}{c}{\textbf{General Agent (GAIA)}} & \textbf{Long Ctx} & \\
\cmidrule(lr){2-4} \cmidrule(lr){5-8} \cmidrule(lr){9-9}
\textbf{Pass@1} & HLE$^*$ & GPQA-D$^{**}$ & LiveBench$^{***}$ & L1 & L2 & L3 & Overall$^\dagger$ & AA-LCR & \textit{Avg~$\Delta$} \\
\midrule
Gemini-3-Flash & 19.40 & 81.31 & 70.00 & 43.40 & 40.70 & 15.38 & 37.58 & 63.00 & \multirow{2}{*}{\textcolor{teal!80!black}{\textbf{+29.24\%}}} \\
\quad+\textsc{DeepTutor} & \textbf{30.80} & \textbf{84.85} & \textbf{96.00} & \textbf{56.60} & \textbf{50.00} & \textbf{23.08} & \textbf{47.88} & \textbf{74.67} & \\
\midrule
Sonnet-4.5 & 8.40 & 72.22 & 64.33 & 37.74 & 30.23 & 7.69 & 29.09 & 53.33 & \multirow{2}{*}{\textcolor{teal!80!black}{\textbf{+32.03\%}}} \\
\quad+\textsc{DeepTutor} & \textbf{14.60} & \textbf{73.23} & \textbf{82.00} & \textbf{50.94} & \textbf{48.84} & \textbf{23.08} & \textbf{45.45} & \textbf{54.00} & \\
\midrule
Qwen-3.5-Plus & 16.80 & 88.38 & 69.00 & 44.23 & 35.71 & 13.64 & 33.94 & 66.00 & \multirow{2}{*}{\textcolor{teal!80!black}{\textbf{+25.69\%}}} \\
\quad+\textsc{DeepTutor} & \textbf{24.20} & 87.88 & \textbf{93.00} & \textbf{50.94} & \textbf{53.49} & \textbf{32.00} & \textbf{49.09} & \textbf{69.67} & \\
\midrule
GPT-5-Mini & 16.46 & 80.81 & 71.00 & 32.08 & 30.23 & 7.69 & 27.27 & 68.67 & \multirow{2}{*}{\textcolor{teal!80!black}{\textbf{+28.51\%}}} \\
\quad+\textsc{DeepTutor} & \textbf{21.20} & 80.30 & \textbf{93.00} & \textbf{54.72} & \textbf{52.33} & \textbf{26.92} & \textbf{49.09} & \textbf{71.00} & \\
\midrule
Minimax-M2.5 & 14.00 & 82.83 & 59.30 & 35.29 & 22.78 & 12.00 & 23.64 & 66.00 & \multirow{2}{*}{\textcolor{teal!80!black}{\textbf{+31.50\%}}} \\
\quad+\textsc{DeepTutor} & \textbf{19.40} & \textbf{83.33} & \textbf{73.00} & \textbf{52.94} & \textbf{44.30} & \textbf{32.00} & \textbf{42.42} & \textbf{76.40} & \\
\bottomrule
\end{tabular}
}
\DTTightTableCaption
\caption{Results of general agentic problem-solving \textbf{Pass@1 scores}.
$^*$We use a fixed 500-question subset.
$^{**}$GPQA-Diamond.
$^{***}$We use the LiveBench reasoning subset.
$^\dagger$GAIA scores are reported \textit{per difficulty level (L1--L3)}.
\textit{Avg~$\Delta$} is defined in \S\ref{sec:generalization}.}
\label{tab:generalization}
\DTTightFloatBottom
\end{table*}

For each backbone model $m$, we summarize performance by averaging the relative gain over its base model across these benchmark groups:
\begin{equation}
\mathrm{Avg}\,\Delta_m =
\frac{1}{|\mathcal{B}|}
\sum_{b \in \mathcal{B}}
\frac{s_{m,b}^{\textsc{DeepTutor}} - s_{m,b}^{\text{base}}}
     {s_{m,b}^{\text{base}}}
\times 100\%,
\end{equation}
where $\mathcal{B}$ denotes the benchmark groups listed above.
GAIA L1--L3 scores are reported only as diagnostic breakdowns.
\DTTightBetween

As shown in Table~\ref{tab:generalization}, the solver-only pipeline improves all five backbone families, with average relative gains ranging from 25.69\% to 32.03\%.
Because personalization, SKG, and DPM are disabled in this setting, the gains point to the general value of the investigate--solve--write scaffold rather than learner-specific context alone.
\DTTightBetween

\section{Related Work}
\label{sec:related}
\DTTightSection
\DTTightParagraph
\paragraph{Personalization and Memory.}
LLM personalization spans dialogue memory via reflective summarization~\citep{tan-etal-2025-prospect} and accumulated user models~\citep{li-etal-2025-hello,cho-etal-2022-personalized}, RAG-style adaptation~\citep{li2025surveypersonalizationragagent,guo2025raganythingallinoneragframework}, workflow induction~\citep{wang2024agentworkflowmemory}, and hierarchical agent memory~\citep{packer2023memgpt,qin2025himem,yang2025gmemory}. Classical learner modeling has long been studied through knowledge tracing, from Bayesian KT~\citep{corbett1994knowledge} to neural and memory-based variants~\citep{piech2015deep,zhang2017dkvmn,ghosh2020akt}. Building on both lines, we propose a tree-structured dynamic memory tailored to tutoring, where interaction traces remain searchable across multiple resolutions.
\DTTightBetween

\DTTightParagraph
\paragraph{AI for Education.}
LLM-based works pursue conversational tutoring with student modeling~\citep{park2024empoweringpersonalizedlearning}, multi-agent pedagogical architectures~\citep{wang2025llmpoweredmultiagentframeworkgoaloriented}, open-source platforms~\citep{xu2025opentutorai}, conversational learning diagnosis~\citep{chen2026parld}, KT-augmented recommendation~\citep{liu2025tutorllm}, and pedagogically aligned tutor optimization~\citep{dinucujianu2025pedagogy}. On the evaluation side, benchmarks such as MathTutorBench~\citep{huang2025mathtutorbench} measure open-ended pedagogical quality, but adaptive tutoring grounded in individualized materials remains underexplored. We unify explanation and question generation through a shared learner model and evaluate the closed loop from a student-centric perspective. Appendix~\ref{app:related} further extends our related-work discussion.

\section{Conclusion}
\DTTightSection
We presented \textsc{DeepTutor}, an agentic personalized tutoring framework that closes the loop between citation-grounded tutoring, difficulty-calibrated question generation, and the Hybrid Personalization Engine.
We also introduced \textsc{TutorBench}, a student-centric benchmark with source-grounded learner profiles and a first-person simulator for interactive evaluation.
Across TutorBench, \textsc{DeepTutor} improves overall quality by 10.8\%, remains stable across five domains, and shows preference trends aligned between human raters and the LLM judge.
Beyond tutoring, the solver-only transfer study yields a 29.4\% average gain across five backbone families, while ablations show that knowledge grounding and learner memory complement each other in anchoring what the tutor says and shaping how it adapts.
The same substrate further supports adaptive learning workflows, interactive books, and proactive multi-channel tutoring agents, suggesting a path from benchmarked tutoring pipelines to long-horizon personalized learning systems.
Together, these results provide a practical foundation for closed-loop personalized tutoring agents.

\section*{Limitations}
\addcontentsline{toc}{section}{Limitations}
\label{sec:limitations}
\DTTightSection
Following recent work, our interactive evaluation relies on LLM-powered student simulators and rubric-based assessors, and therefore inherits the usual gap between controlled simulation and real learner behavior. A large-scale validation with human students could be further considered.
Like most agentic systems, the multi-stage pipeline trades additional inference cost for stronger controllability and personalization.
Moreover, the paradigm of TutorBench could be scaled up to finer-grained courses, longer curricular trajectories, and larger learner populations.
The system extensions described in Section~\ref{sec:extensions}, including the Book Engine and Partners, are presented as architectural instantiations; their effects on retention, engagement, interruption cost, and real learner outcomes require longitudinal human studies.

\paragraph{Broader impacts.}
Personalized tutoring systems may broaden access to adaptive educational support, especially when high-quality human tutoring is scarce. At the same time, any LLM-based tutor can inherit general LLM risks such as hallucinated facts, overconfident explanations, or subtle reasoning errors. Real-world use should therefore treat generated tutoring content as assistive rather than authoritative, preserve source-grounded verification, and encourage students and educators to check important claims against trusted course materials or human instructors.

\bibliographystyle{abbrvnat}
\nobibliography*

\begingroup
\hypersetup{urlcolor=black}
\bibliography{custom}
\endgroup

\appendix
\DTAppendixContentsPage
% =============================================================================
% Appendix
% =============================================================================
\section{Extended Related Work}
\label{app:related}
% ============================================================

This appendix expands the literature review in \S\ref{sec:related} that provides context for the design of an agentic tutoring system, including agentic tool use, memory, learner modeling, and LLM-based educational agents.

% ------------------------------------------------------------
\subsection{Tool-Augmented LLM Agents}
\label{app:related:agents}
% ------------------------------------------------------------

Equipping language models with external tools has progressed from
\emph{static} integration to \emph{dynamic} reasoning.
Toolformer~\citep{schick2024toolformer} showed that LLMs can learn to
insert API calls through self-supervised fine-tuning.
ReAct~\citep{yao2023reactsynergizingreasoningacting} introduced a more
flexible paradigm by interleaving chain-of-thought reasoning with
environment-facing actions, significantly improving multi-step retrieval
and decision-making.
Subsequent work has expanded planning capabilities in various
directions~\citep{zhao2025llmbasedagenticreasoningframeworks,
wu-etal-2025-agentic, li2025intheflowagenticoptimizationeffective},
while long-horizon agents such as
IterResearch~\citep{chen2026iterresearch} tackle coherent state
maintenance across many reasoning steps.
Many existing agentic frameworks are still organized around
\emph{bounded} episodes, where a single query triggers a reasoning
trajectory that ends once an answer is produced.
Tutoring, however, often extends across multiple sessions and benefits
from maintaining continuity with an evolving model of the individual
learner.

% ------------------------------------------------------------
\subsection{LLM Personalization and Memory}
\label{app:related:memory}
% ------------------------------------------------------------

This stateful requirement connects agentic tool use to a parallel line
of work on LLM personalization and memory, which has been pursued along
three main axes.
\emph{Dialogue-level memory} maintains context across conversation turns
through reflective
summarization~\citep{tan-etal-2025-prospect}, accumulated user
models~\citep{li-etal-2025-hello}, or implicit persona
detection~\citep{cho-etal-2022-personalized}.
\emph{Workflow-level memory} goes beyond individual episodes: Agent
Workflow Memory~\citep{wang2024agentworkflowmemory} induces reusable
task workflows from past experience for transfer across structurally
similar future tasks.
\emph{RAG-based personalization} adapts the retrieval pipeline itself to
user preferences~\citep{li2025surveypersonalizationragagent}, with
frameworks such as
RAG-Anything~\citep{guo2025raganythingallinoneragframework} providing
flexible multimodal retrieval backbones.
MemGPT~\citep{packer2023memgpt} offers a complementary perspective,
managing unbounded context through OS-inspired memory paging.
More recently, hierarchical memory architectures have emerged:
HiMem~\citep{qin2025himem} segments long-horizon dialogues into
episode and note memories with conflict-aware reconsolidation, while
G-Memory~\citep{yang2025gmemory} organizes multi-agent collaboration
traces into a three-tier graph hierarchy of insights, queries, and
interactions, achieving up to 20\% gains in embodied-action benchmarks.
These approaches provide useful building blocks. For tutoring settings,
it is also useful to organize traces in domain-specific, typed, and
semantically searchable forms across multiple resolutions (sessions,
plan steps, tool invocations), so that downstream agents can reuse them
for both explanation and question generation.

% ------------------------------------------------------------
\subsection{Knowledge Tracing and Intelligent Tutoring Systems}
\label{app:related:kt}
% ------------------------------------------------------------

The need for structured learner modeling has a long history in education.
Classical \emph{knowledge tracing} (KT) estimates per-skill mastery
probabilities from interaction logs: Bayesian Knowledge
Tracing~\citep{corbett1994knowledge} uses two-state hidden Markov
models, Deep Knowledge Tracing~\citep{piech2015deep} replaces
hand-crafted features with recurrent networks, DKVMN~\citep{zhang2017dkvmn}
separates concept keys from evolving mastery values in a key-value memory,
and context-aware attention models~\citep{ghosh2020akt} further improve
interpretability through monotonic attention over learner histories.
Early \emph{Intelligent Tutoring Systems} (ITS) such as
Cognitive Tutor~\citep{anderson1995cognitive} were already providing
step-level guidance through domain-specific production rules, enabling
fine-grained diagnosis of learner actions at the cost of substantial
manual domain engineering.
Recent work points to a growing shift toward LLM-based tutoring, which
brings broad domain coverage and flexible natural-language interaction
while often placing less emphasis on explicit, inspectable learner
models of the kind common in classical
ITS~\citep{letourneau2025systematicreviewaidriven,chu-etal-2025-llm}.

% ------------------------------------------------------------
\subsection{LLM-Based Educational Agents}
\label{app:related:edu}
% ------------------------------------------------------------

Building on this shift, recent work has explored LLMs in dedicated
tutoring roles along several complementary directions.
\citet{park2024empoweringpersonalizedlearning} develop a conversational
system with explicit student modeling, showing that tracking student
state improves learning outcomes.
\citet{wang2025llmpoweredmultiagentframeworkgoaloriented} decompose the
tutoring task across specialized agents for curriculum planning,
dialogue, and assessment.
\citet{frankford2024aitutoringsoftwareengineering} apply LLM tutoring to
software engineering education, suggesting that richer student models
can support higher-quality feedback.
Open TutorAI~\citep{xu2025opentutorai} provides an open-source platform
with customizable avatars for multimodal interaction, while
ParLD~\citep{chen2026parld} introduces a preview--analyze--reason
pipeline for conversational learning diagnosis via multi-agent
collaboration.
TutorLLM~\citep{liu2025tutorllm} couples knowledge tracing with
RAG-based recommendation, and \citet{dinucujianu2025pedagogy} align
tutoring LLMs with pedagogical principles via reinforcement learning
from simulated student interactions.
Taken together, this literature now spans feedback generation,
curriculum design, adaptive assessment, pedagogically aligned training,
and open-ended evaluation~\citep{chu-etal-2025-llm,huang2025mathtutorbench}.
Across these systems, problem tutoring and question generation are often
treated as separate components, with limited sharing of learner
representations or performance signals. This design space is related to
the bidirectional task coupling described in \S\ref{sec:system}. At the
same time, the literature remains broad in scope, and several closely
related systems do not yet offer end-to-end reproducible releases for
our exact tutoring setting.

% ============================================================
\section{Implementation Details of DeepTutor System}
\label{app:prompts}
% ============================================================

This appendix summarizes the two system-level loops behind \textsc{DeepTutor}:
the closed-loop tutoring cycle (\S\ref{app:algorithm}) and the TutorBot
autonomous agent loop (\S\ref{app:tutorbot}).

% ------------------------------------------------------------
\subsection{Closed-Loop Personalized Tutoring}
\label{app:algorithm}
% ------------------------------------------------------------

Algorithm~\ref{alg:loop} gives the session-level execution order of
\textsc{DeepTutor}'s closed tutoring cycle. Starting from an initial
learner profile $\mathcal{D}^{(0)}$ and an empty trace forest
$\mathcal{F}^{(0)}$, each new session first assembles a personalized
memory context $\mathcal{C}_{\text{mem}}$ from prior interactions and
profile excerpts. The algorithm then branches according to the task type:
problem tutoring uses retrieval-grounded investigation, planning,
step-by-step tool use, and iterative answer writing, while question
generation uses personalized idea filtering followed by critic-guided
question, answer, and explanation construction. In both branches, the completed interaction is converted
into a trace tree and fed back into the memory update, so subsequent
sessions can reuse both high-level profile summaries and fine-grained
interaction evidence.

\begin{algorithm}[ht!]
\caption{Closed-Loop Personalized Tutoring}\label{alg:loop}
\small
\begin{algorithmic}[1]
\Require learner profile $\mathcal{D}^{(0)}$;\; trace forest $\mathcal{F}^{(0)} = \emptyset$
\For{session $j = 1, 2, \ldots$}
    \State $\mathcal{C}_{\text{mem}} \gets \Call{ProfileInject}{\mathcal{D}^{(j{-}1)}, \mathcal{F}^{(j{-}1)}, \text{task}_j}$
    \If{$\text{task}_j = (\textsc{Solve},\, q)$}
        \State $\mathcal{C}_{\text{rag}} \gets \textsc{RagTool}(q)$
        \State $\mathcal{E}_{\text{plan}} \gets \Call{Investigate}{q,\;\mathcal{C}_{\text{rag}},\;\mathcal{C}_{\text{mem}},\;\mathcal{F}^{(j{-}1)}}$ \Comment{\ding{172}}
        \State $\mathcal{P}{=}\langle s_1,\dots,s_K\rangle \gets \Call{Plan}{q,\;\mathcal{E}_{\text{plan}},\;\mathcal{C}_{\text{mem}}}$
        \For{$k = 1$ \textbf{to} $K$} \Comment{\ding{173}}
            \While{$s_k$ not resolved}
                \State $(\alpha_t,\, \mathbf{s}_t) \gets \Call{ToolAgent}{q, \mathcal{P}, \mathcal{S}, \mathcal{C}_{\text{rag}}, \mathcal{C}_{\text{mem}}}$
                \State $\mathcal{S} \gets \mathcal{S} \cup \{\mathbf{s}_t\}$
                \If{$\alpha_t = \textsc{Replan}$}
                    \State $\mathcal{P} \gets \Call{Revise}{\mathcal{P}, \mathcal{S}}$
                \EndIf
            \EndWhile
            \State $\mathcal{S} \gets \Call{Compress}{\mathcal{S}, k}$
        \EndFor
        \State $a \gets \Call{IterativeWrite}{\mathcal{S},\;\mathcal{C}_{\text{mem}}}$ \Comment{\ding{174}}
        \State $T_j \gets \Call{BuildTrace}{q, a}$
    \Else \Comment{$\text{task}_j = (\textsc{Generate},\, \tau)$}
        \State $\mathcal{C}_{\text{rag}} \gets \textsc{RagTool}(\tau)$
        \State $f_{\text{idea}} \gets \emptyset$
        \Repeat
            \State $\mathcal{I} \gets \Call{IdeaAgent}{\tau,\;\mathcal{C}_{\text{rag}},\;\mathcal{C}_{\text{mem}},\;f_{\text{idea}}}$ \Comment{\ding{175}}
            \State $(\textit{continue},\, f_{\text{idea}},\, \mathcal{T}_{1..N}) \gets \Call{EvaluateIdeas}{\mathcal{I},\;\tau,\;\mathcal{C}_{\text{mem}}}$
        \Until{$\lnot\,\textit{continue}$}
        \For{$i = 1$ \textbf{to} $N$} \Comment{\ding{176}}
            \Repeat
                \State $(q_i, a_i, e_i) \gets \Call{Generator}{\mathcal{T}_i,\;\mathcal{C}_{\text{rag}},\;\mathcal{C}_{\text{mem}}}$
                \State $(\textit{pass}, f_i) \gets \Call{Validator}{q_i, a_i, e_i, \mathcal{T}_i}$
                \If{$\lnot\,\textit{pass}$} $\mathcal{T}_i \gets \Call{Revise}{\mathcal{T}_i, f_i}$
                \EndIf
            \Until{\textit{pass}}
        \EndFor
        \State $T_j \gets \Call{BuildTrace}{\tau, \{(q_i, a_i, e_i)\}_{i=1}^{N}}$
    \EndIf
    \State $\mathcal{F}^{(j)} \gets \mathcal{F}^{(j{-}1)} \cup \{T_j\}$
    \State $\mathcal{D}^{(j)} \gets \Call{MemoryUpdate}{\mathcal{D}^{(j{-}1)}, T_j}$
\EndFor
\end{algorithmic}
\end{algorithm}

% ------------------------------------------------------------

% ------------------------------------------------------------
\subsection{TutorBot Autonomous Agent Loop}
\label{app:tutorbot}
% ------------------------------------------------------------

Algorithm~\ref{alg:tutorbot} summarizes the autonomous agent loop used by
\textsc{TutorBot}, the runtime behind the Partners multi-channel deployment
layer described in \S\ref{sec:extensions:tutorbot}. Each inbound channel message triggers a
three-phase cycle: context assembly, tool-augmented reasoning, and memory
persistence. The bot first composes persona, long-term memory, active skills,
and recent history into a working context; it then runs a bounded ReAct-style
tool-calling loop; finally, it persists the turn and consolidates older
messages into durable memory when the session approaches the context budget.

\begin{algorithm}[ht!]
\caption{TutorBot Autonomous Agent Loop}\label{alg:tutorbot}
\small
\begin{algorithmic}[1]
\Require message bus $\mathcal{B}$;\; LLM provider $\mathcal{M}$;\; tool registry $\mathcal{T}$;\; session store $\mathcal{S}$;\; context window $W$;\; max iterations $I_{\max}$
\Statex
\Loop \Comment{Main event loop}
    \State $m \gets \mathcal{B}.\textsc{ConsumeInbound}()$ \Comment{Block until next message}
    \State $\sigma \gets \mathcal{S}.\textsc{GetOrCreate}(m.\textit{session\_key})$
    \Statex
    \Comment{\textbf{Phase 1: Context Assembly}}
    \State $\textit{sys} \gets \textsc{BuildContext}(\sigma.\textit{soul},\; \sigma.\textit{memory},\; \sigma.\textit{skills})$
    \State $H \gets \sigma.\textsc{GetHistory}()$ \Comment{Unconsolidated messages}
    \State $\textit{msgs} \gets [\textit{sys}] \,\|\, H \,\|\, [\textsc{Wrap}(m)]$
    \Statex
    \Comment{\textbf{Phase 2: ReAct Tool-Calling Loop}}
    \For{$i = 1$ \textbf{to} $I_{\max}$}
        \State $r \gets \mathcal{M}.\textsc{Chat}(\textit{msgs},\; \mathcal{T}.\textsc{Definitions}())$
        \If{$r$ has no tool calls}
            \State $\textit{response} \gets r.\textit{content}$ \Comment{Final answer}
            \State \textbf{break}
        \EndIf
        \State Append assistant message (with tool calls) to $\textit{msgs}$
        \ForAll{tool call $\tau$ in $r.\textit{tool\_calls}$}
            \State $\textit{result} \gets \mathcal{T}.\textsc{Execute}(\tau.\textit{name},\; \tau.\textit{args})$
            \State Append tool result to $\textit{msgs}$
        \EndFor
    \EndFor
    \Statex
    \Comment{\textbf{Phase 3: Persist \& Consolidate}}
    \State $\sigma.\textsc{AppendTurn}(\textit{msgs})$
    \If{$\textsc{EstimateTokens}(\sigma) > W$} \Comment{Context pressure}
        \State $\textit{chunk} \gets \sigma.\textit{messages}[\sigma.\textit{ptr} : \textit{boundary}]$
        \State $\textsc{Consolidate}(\textit{chunk},\; \sigma.\textit{memory})$ \Comment{LLM-based distillation}
        \State $\sigma.\textit{ptr} \gets \textit{boundary}$
    \EndIf
    \State $\mathcal{S}.\textsc{Save}(\sigma)$
    \State $\mathcal{B}.\textsc{PublishOutbound}(\textit{response},\; m.\textit{channel})$
\EndLoop
\end{algorithmic}
\end{algorithm}

The \textsc{BuildContext} procedure concatenates the bot's identity,
runtime metadata, persona specification, user preferences, memory summaries,
and active skill descriptions. The \textsc{Consolidate} procedure uses an
LLM-based memory consolidator to append a timestamped history entry while
updating the long-term learner profile. The concrete prompt templates and tool
schemas are released with the repository rather than enumerated here.

\section{TutorBench Construction Details}
\label{app:tutorbench}
% ============================================================

This appendix supplements \S\ref{sec:tutorbench} with full
construction details.
Each benchmark \emph{entry} combines a learner profile, a set of
knowledge gaps, an interactive task, and the supporting source content.
Below we describe the source PDF inventory
(\S\ref{app:tutorbench:sources}), the rejection-sampling procedure
that produces high-quality tasks (\S\ref{app:tutorbench:task}), and
the relationship between Trace Forest and benchmark-time simulation
(\S\ref{app:tutorbench:traceforest}).

% ------------------------------------------------------------
\subsection{PDF Source Inventory}
\label{app:tutorbench:sources}
% ------------------------------------------------------------

TutorBench is built from 30 KBs derived from source PDFs. For textbook
sources, each book is partitioned into three chapter-contiguous PDFs
before KB construction; for research sources, each paper forms a single
KB. The resulting PDF inventory is listed below.

\paragraph{Textbook-derived PDFs.}
\begin{itemize}[leftmargin=1.5em,itemsep=1pt,topsep=2pt]
  \item All textbook PDFs are sourced from OpenStax.
  \item \textit{Calculus Volume 2} (OpenStax, \url{https://openstax.org/details/books/calculus-volume-2/}): Chapters 1--2, Chapters 3--4, and Chapters 5--6.
  \item \textit{Calculus Volume 3} (OpenStax, \url{https://openstax.org/details/books/calculus-volume-3/}): Chapters 1--2, Chapters 3--4, and Chapters 5--6.
  \item \textit{Principles of Economics 3e} (OpenStax, \url{https://openstax.org/details/books/principles-economics-3e/}): Chapters 1--6, Chapters 7--12, and Chapters 13--18.
  \item \textit{Foundations of Information Systems} (OpenStax, \url{https://openstax.org/details/books/foundations-information-systems}): Chapters 1--3, Chapters 4--6, and Chapters 7--9.
  \item \textit{Introduction to Computer Science} (OpenStax, \url{https://openstax.org/details/books/introduction-computer-science}): Chapters 1--3, Chapters 4--6, and Chapters 7--9.
  \item \textit{Introduction to Philosophy} (OpenStax, \url{https://openstax.org/details/books/introduction-philosophy}): Chapters 1--4, Chapters 5--8, and Chapters 9--12.
  \item \textit{Introduction to Business} (OpenStax, \url{https://openstax.org/details/books/introduction-business}): Chapters 1--5, Chapters 6--10, and Chapters 11--15.
  \item \textit{Writing Guide with Handbook} (OpenStax, \url{https://openstax.org/details/books/writing-guide}): Chapters 1--6, Chapters 7--12, and Chapters 13--18.
\end{itemize}

\paragraph{Paper-derived PDFs.}
\begin{itemize}[leftmargin=1.5em,itemsep=1pt,topsep=2pt]
  \item \textit{Memory in the Age of AI Agents}~\citep{hu2025memory}.
  \item \textit{DeepSeek-R1 incentivizes reasoning in LLMs through reinforcement learning}~\citep{guo2025deepseekr1}.
  \item \textit{LiveCodeBench: Holistic and Contamination Free Evaluation of Large Language Models for Code}~\citep{jain2025livecodebench}.
  \item \textit{OpenVLA: An Open-Source Vision-Language-Action Model}~\citep{kim2025openvla}.
  \item \textit{Towards Reasoning Era: A Survey of Long Chain-of-Thought for Reasoning Large Language Models}~\citep{chen2025towardsreasoningera}.
  \item \textit{YOLOv9: Learning What You Want to Learn Using Programmable Gradient Information}~\citep{wang2024yolov9}.
\end{itemize}

\subsection{Task Generation and Rejection Sampling}
\label{app:tutorbench:task}
% ------------------------------------------------------------

Given a profile $p$ and its associated knowledge gaps $G$, the task
generator creates interactive learning tasks that naturally expose
one or more gaps during a tutoring conversation.
Four task types are supported---\emph{concept understanding} (30\%),
\emph{problem solving} (30\%), \emph{application} (20\%), and
\emph{comparison} (20\%)---with sampling weights reflecting typical
student learning request distributions.
A batch rejection sampler then filters out tasks whose gaps are
incoherent, misaligned with the task description, or conversationally
unnatural / internally inconsistent,
and regenerates until the minimum task count is satisfied.
Algorithm~\ref{alg:task} gives the full procedure.

\begin{algorithm}[ht!]
\caption{Task Generation with Rejection Sampling}\label{alg:task}
\small
\begin{algorithmic}[1]
\Require Profile $p$, gaps $G$, knowledge scope $\text{scope}$,
         source pages $S$; minimum tasks $T_{\min} = 3$
\State $\text{accepted} \gets \emptyset$
\Repeat
    \State $\mathcal{T}^{\text{cand}} \gets
           \textsc{GenerateTasks}(p,\, G,\, \text{scope})$
    \Comment{batch LLM generation}
    \For{each candidate task $t \in \mathcal{T}^{\text{cand}}$}
        \State $\text{ok} \gets \textsc{RejectSample}(t,\, G,\, S)$
        \Comment{3-criterion check (below)}
        \If{$\text{ok}$}
            \State $\text{accepted} \gets \text{accepted} \cup \{t\}$
        \EndIf
    \EndFor
\Until{$|\text{accepted}| \geq T_{\min}$}
\State \Return $\text{accepted}$
\end{algorithmic}
\end{algorithm}

\noindent The \textsc{RejectSample} oracle verifies three criteria:
\begin{enumerate}[nosep,leftmargin=*]
  \item \textbf{Gap coherence.}
        The gaps targeted by the task must be logically related and
        form a coherent learning topic; tasks requiring the student to
        jump between disconnected concepts are rejected.
  \item \textbf{Task--gap fit.}
        The task description, initial student message, and success
        criteria must organically connect to the targeted gaps---pursuing
        the task should naturally expose those gaps without artificial
        steering.
  \item \textbf{Conversational naturalness and consistency.}
        The resulting task should read like a plausible learner request
        for the given profile, rather than an artifact assembled solely
        to surface benchmark gaps. Items whose gaps contradict one
        another or conflict with the task description are rejected under
        this criterion.
\end{enumerate}
The oracle is instructed to apply strict criteria: when in doubt, reject.
This conservative stance accepts a higher regeneration rate in exchange
for higher task quality, a trade-off justified by the one-time offline
nature of benchmark construction.
Each accepted task carries a structured record: task type, title,
description, initial student message (matched to the profile's
personality), targeted gap IDs, expected gap exposure trajectory,
success criteria, and difficulty rating. Step~1 then packages the
profile, selected gaps, accepted task, and the corresponding sampled
\texttt{source\_content} into one benchmark entry, which becomes the
unit consumed by the simulation and evaluation pipeline.

% ------------------------------------------------------------
\subsection{Trace Forest and Benchmark Runtime Memory}
\label{app:tutorbench:traceforest}
% ------------------------------------------------------------

The benchmark entry files produced during construction do \emph{not}
serialize a Trace Forest.
Trace Forest is a \emph{runtime} memory mechanism: during Step~2
simulation, \textsc{DeepTutor} backends with cross-session memory
enabled build and query a Trace Forest to condition both
problem-tutoring responses and practice-question generation.
Ablated variants disable this memory while keeping the underlying
benchmark entries unchanged.
The key implication for reproducibility is that the benchmark fixes
the same entry objects for all systems; only the \textsc{DeepTutor}
family differs in whether it reads from and writes to this runtime
memory during multi-session simulation.

% ============================================================
\section{Evaluation Protocol and Hyperparameters}
\label{app:eval}
% ============================================================

This section provides full evaluation details for the interactive evaluation (\S\ref{app:eval:interactive}), the human preference alignment study (\S\ref{app:eval:human_alignment}), and the general problem-solving benchmarks (\S\ref{app:eval:general}), together with baseline definitions (\S\ref{app:eval:baselines}), hyperparameters (\S\ref{app:eval:hparams}), and implementation specifics (\S\ref{app:eval:impl}).

% ------------------------------------------------------------
\subsection{First-Person Interactive Evaluation Details}
\label{app:eval:interactive}
% ------------------------------------------------------------

Figure~\ref{fig:eval_protocol} in the main paper summarizes the full
interactive evaluation loop used for TutorBench; this appendix provides the
implementation details.
Each evaluation begins from a benchmark entry containing a learner profile,
diagnosed knowledge gaps, an interactive task, and relevant source materials.
The student simulator turns these inputs into first-person beliefs and
maintains a dynamic tutoring list during the dialogue.
At each turn, it either asks the tutor for help on an unfinished tutoring item
or, once the tutoring phase is complete, asks the tutor to generate a customized
quiz.
All tutor responses are recorded as traces and later judged with personalized
rubrics derived from the same learner state and task context.

\subsubsection{Student Simulator}
\label{app:eval:simulator}

The student simulator is an LLM agent initialized from a TutorBench
entry (profile + gaps + task).
Each gap is pre-transformed into a first-person belief statement so
the simulator \emph{acts as} the student rather than narrating errors
from an outside view.
The system prompt below governs simulator behavior throughout the
interaction; the template variables are filled from the entry at
initialization time.

\begin{lstlisting}[caption={Student Simulator System Prompt}]
You are role-playing as a student seeking help from an AI tutor.
Stay in character throughout the entire conversation. Never break
character.

# Who You Are
{personality}

# Your Background
{education_background}

# Why You Are Here
{learning_purpose}

# What You Know
You are confident about these topics:
{known_well}

You have vague or partial understanding (you might get details wrong):
{partially_known}

You have NO knowledge of the following (never encountered):
{unknown}

# What You Believe (IMPORTANT -- these feel true to you)
{beliefs}

# Behavioral Rules
1. NEVER display knowledge listed as "unknown". If the tutor asks
   about something you don't know, say so honestly.
2. When the tutor teaches something new, try to rephrase it in your
   own words. It's okay to rephrase imperfectly.
3. If the tutor asks "do you understand?", be honest based on whether
   the explanation actually addressed your confusion.
4. Keep responses concise: 1-4 sentences typically.
5. Stay consistent with your personality throughout.
6. You may ask follow-up questions or request examples.
7. When you feel you've understood, PREFER asking for a practice
   problem before ending -- testing yourself helps solidify learning.

# Ending the conversation (ONLY when done)
Use [ACTION: task_complete] ONLY if ALL conditions are true:
  - You are explicitly done.
  - You have zero remaining questions.
  - You are NOT requesting anything else.
  - Your message is a natural closing/goodbye.

CRITICAL: If your message contains any question, uncertainty, or
request for more help, do NOT use [ACTION: task_complete].
\end{lstlisting}

\noindent \textbf{Dynamic gap pacing.}
The \texttt{\{beliefs\}} placeholder is populated with first-person
formulations of each knowledge gap (e.g., ``I think the Poynting
vector always points in the direction of wave propagation'').
The simulator's behavioral rules enforce realistic resistance patterns:
the student may defend misconceptions, partially accept corrections,
and only signal completion after sufficient scaffolded explanation.
This prevents trivially short sessions where the student immediately
agrees with any tutor statement.

\subsubsection{Evaluation Rubrics}
\label{app:eval:rubrics}

Tutoring quality is assessed along ten dimensions, each scored on a
1--5 Likert scale by an LLM judge.
The ten metrics are organized into two groups---five
\textbf{tutoring-side} metrics that evaluate every tutor turn during the
dialogue, and five \textbf{practice-side} metrics that score the
generated questions after each session.
In addition, the evaluator records raw turn counts as a non-LLM
structural statistic.
The pipeline reports all metrics independently; no weighted composite
score is computed.
The dimensions operationalize constructs from source-grounded
generation, learner modeling and formative feedback, multiple
representations for explanation, and educational question
assessment~\citep{maynez2020faithfulness,shute2008focus,
corbett1994knowledge,ainsworth2006deft,kurdi2020systematic}.

\paragraph{Tutoring-Side Metrics (turn-level, 1--5 each).}
Each metric below is evaluated on every student-to-tutor turn pair.
When practice questions are present, the final dialogue turn that emits
them is excluded from tutoring-side scoring to keep the tutoring and
question-generation phases separate.

\begin{itemize}[nosep,leftmargin=*]
\item \textbf{Source Faithfulness~(SF)} captures factual consistency
  with the source material provided in each benchmark entry, following
  work on faithful and attributable generation~\citep{maynez2020faithfulness,
  rashkin2023measuring}.
  The judge examines four sub-dimensions:
  (a)~\emph{factual consistency}---whether tutor claims agree with the
  source rather than contradicting it;
  (b)~\emph{terminological consistency}---whether the tutor follows the
  source's definitions and framing;
  (c)~\emph{source grounding}---whether the response shows evidence of
  drawing on source-specific content, not just generic knowledge; and
  (d)~\emph{source attribution}---whether the tutor explicitly
  attributes information via citation markers or natural-language
  references.
  Pedagogical elaboration beyond the source---analogies, worked
  examples, motivating applications---is explicitly \emph{not}
  penalized, as long as it does not distort the source.
  A score of~5 requires accuracy, consistent terminology, clear source
  grounding, and explicit attribution; a score of~1 indicates core
  claims that are wrong or that directly conflict with the source.

\item \textbf{Personalization~(PER)} reflects how specifically the
  tutor addresses the learner's current confusion while remaining
  aligned with the learner state developed over the session~\citep{
  vygotsky1978mind,corbett1994knowledge}.
  The judge checks whether the response is tailored to the student's
  profile, the misconceptions surfaced in prior turns, and the
  particular question just asked---rather than offering a generic
  textbook answer.
  A score of~5 requires strong, visible tailoring to this specific
  learner's misunderstanding and conversational trajectory;
  a score of~3 is partially tailored but still generic in key parts;
  a score of~1 indicates a fully generic response that ignores
  learner-specific context entirely.

\item \textbf{Applicability~(APP)} evaluates whether the guidance
  is concrete and immediately actionable for the student~\citep{
  hattie2007power,shute2008focus}.
  High scores require the tutor to provide specific, executable content
  such as worked examples, step-by-step procedures, verifiable
  calculations, or diagnostic questions that let the student make
  immediate progress on their problem.
  A score of~5 means the student can directly act on the
  instruction; a score of~1 means the response is vague, abstract, or
  impossible to verify.

\item \textbf{Vividness~(VID)} rewards representationally rich,
  engaging presentation that goes beyond monolithic plain-text
  prose~\citep{ainsworth2006deft,mayer2003promise}.
  The judge assesses modality diversity: effective use of structured
  lists, equations paired with intuitive explanations, comparison
  tables, code demonstrations, or other representational forms that
  genuinely aid understanding.
  A score of~5 indicates rich multimodal delivery whose format choices
  clearly enhance learning; a score of~1 indicates flat, undifferentiated
  text with no meaningful modality variation.

\item \textbf{Logical Depth~(LD)} assesses the presence of explicit
  multi-step causal or analytical reasoning, as opposed to bare
  assertions~\citep{chi1989self,koedinger2012knowledge}.
  The judge looks for coherent reasoning chains: does the tutor explain
  \emph{why} a result holds, trace cause-and-effect relationships, or
  walk the student through intermediate steps?
  A score of~5 requires clear, well-structured multi-step reasoning;
  a score of~1 means the tutor provides conclusions or facts without
  any substantive justification.
\end{itemize}

\paragraph{Practice-Side Metrics (session-level, 1--5 each).}
After the main tutoring dialogue concludes, the tutor generates five
practice multiple-choice questions (MCQs).
Each question is evaluated independently across the following five
dimensions:

\begin{itemize}[nosep,leftmargin=*]
\item \textbf{Fitness~(FIT)} measures alignment with the
  session-level diagnosed weaknesses at appropriate difficulty~\citep{
  wood1976role,lord1980applications}.
  The judge checks whether the question targets the knowledge gaps
  surfaced during the dialogue, matches the student's proficiency
  level, and is neither trivially easy nor unreasonably hard given the
  learner profile.
  A score of~5 means the question precisely targets the student's weak
  points at a suitable challenge level; a score of~1 means the
  question is irrelevant to the diagnosed gaps or grossly
  miscalibrated in difficulty.

\item \textbf{Groundedness~(GND)} requires factual anchoring in the
  source material, mirroring the structure of Source Faithfulness on
  the solve side~\citep{lewis2020retrieval,gao2023enabling}.
  The judge checks four parallel sub-dimensions: factual consistency,
  terminological consistency, source grounding, and source attribution
  of the question stem, correct answer, and explanation.
  A score of~5 means the question content is fully consistent with and
  clearly derived from the source; a score of~1 means the question
  uses fabricated claims or definitions that conflict with the source.

\item \textbf{Diversity~(DIV)} rewards novelty in angle and cognitive
  demand~\citep{anderson2001taxonomy,kurdi2020systematic}.
  The judge evaluates whether the question presents a distinct
  perspective, question format, or cognitive action (e.g., application,
  analysis, comparison) compared with typical rote recall items.
  A score of~5 reflects a creative, non-templated question that
  engages higher-order thinking; a score of~1 indicates a standard
  fill-in-the-blank or direct-recall item indistinguishable from
  routine exercises.

\item \textbf{Answer Quality~(ANS)} evaluates the correctness of the
  reference answer and the plausibility of distractors~\citep{
  haladyna2002review}.
  The judge verifies that the correct option is unambiguously right,
  that the explanation is clear and consistent with the chosen answer,
  and that distractors are wrong yet plausible enough to be
  pedagogically informative.
  A score of~5 requires a flawless key, a lucid explanation, and
  well-crafted distractors; a score of~1 indicates an incorrect key,
  a contradictory explanation, or nonsensical options.

\item \textbf{Cross Concept~(CC)} tests whether the question
  meaningfully integrates multiple concepts surfaced throughout the
  session, rather than testing a single isolated fact~\citep{
  chi1981categorization,barnett2002when}.
  The judge checks whether the question requires the student to
  synthesize ideas from different parts of the material or to reason
  about relationships between concepts.
  A score of~5 means the question demands genuine cross-concept
  reasoning; a score of~1 means it tests only one narrow, isolated
  point.
\end{itemize}

\paragraph{Auxiliary statistics.}
Beyond the ten judge-scored metrics, the evaluator records raw
turn counts (student turns, tutor turns, and paired turns) as a
structural measure of dialogue length. These counts are reported
directly and are not normalized to the 1--5 scale.
Additionally, the evaluator tracks judge reliability statistics such as
parse-failure rates per metric; these are used to monitor evaluation
stability but are not part of the quality scores themselves.

\subsubsection{Protocol Details}
\label{app:eval:protocol}

The full benchmark pipeline runs in three stages.
\textbf{Step~1} (entry generation): TutorBench entries are generated
offline using the pipeline described in Appendix~\ref{app:tutorbench}.
\textbf{Step~2} (multi-session simulation): for each
KB/profile/backend tuple, the student simulator is initialized from the
entry set and runs up to \texttt{max\_turns}~=~30 student turns per
session in the default batch pipeline; transcripts are saved together
with five practice questions produced at session end.
\textbf{Step~3} (LLM-as-judge): each saved transcript is scored for
source faithfulness, four teaching-quality dimensions, turn counts, and
practice-question quality.

For the reported interactive evaluation, \textit{Gemini-3-Flash}
powers both the student simulator and all tutor backbones, including
\textsc{DeepTutor}, while \textit{Claude Sonnet~4.6} serves as the fixed
judge with temperature~=~0.0. To mitigate judge variance, each
transcript is scored three times and the scores are averaged.

All 1--5 scores are averaged to four decimal places at session,
multi-session, and backend-summary levels. The current pipeline does
\emph{not} compute a weighted overall benchmark score; instead it reports
the independent metrics directly.

\subsubsection{Cross-Domain Metric Table}
\label{app:results:breakdown}

Table~\ref{tab:domain} reports the complete domain-by-metric breakdown
for the first-person interactive simulation visualized in
Figure~\ref{fig:domain}.

\begin{table*}[ht!]
\centering
\scriptsize
\setlength{\tabcolsep}{4pt}
\resizebox{\textwidth}{!}{%
\begin{tabular}{l ccccc c ccccc c c}
\toprule
& \multicolumn{6}{c}{\textbf{Tutoring Quality}}
& \multicolumn{6}{c}{\textbf{Practice Quality}}
& \\
\cmidrule(lr){2-7} \cmidrule(lr){8-13}
\textbf{Domain}
  & SF & PER & APP & VID & LD & \textit{Avg}
  & FIT & GND & DIV & ANS & CC & \textit{Avg}
  & OQ \\
\midrule
Humanities
  & 3.24 & 4.55 & 4.45 & 4.75 & 4.71 & 4.34
  & 3.15 & 2.80 & 3.58 & 3.82 & 3.32 & 3.33
  & 3.84 \\
Sciences
  & 3.56 & 4.64 & 4.63 & 4.80 & 4.67 & 4.46
  & 3.43 & 3.03 & 3.16 & 4.14 & 3.53 & 3.46
  & 3.96 \\
Engineering
  & 3.59 & 4.54 & 4.48 & 4.66 & 4.29 & 4.31
  & 3.24 & 3.05 & 3.53 & 4.04 & 3.47 & 3.46
  & 3.89 \\
Business
  & 3.53 & 4.56 & 4.74 & 4.93 & 4.73 & 4.50
  & 3.54 & 2.92 & 3.42 & 4.16 & 3.39 & 3.48
  & 3.99 \\
Research
  & 2.87 & 4.64 & 4.52 & 4.92 & 4.64 & 4.32
  & 3.39 & 3.02 & 3.52 & 3.75 & 3.21 & 3.38
  & 3.85 \\
\bottomrule
\end{tabular}
}
\caption{Full cross-domain results for the first-person interactive simulation.
  The main paper visualizes the same data in Figure~\ref{fig:domain}.
  OQ is the average over the five tutoring-side and five practice-side metrics.}
\label{tab:domain}
\end{table*}

% ------------------------------------------------------------
\subsection{Baseline and Variant Definitions}
\label{app:eval:baselines}
% ------------------------------------------------------------

We compare \textsc{DeepTutor} against four prompt-based tutor baselines
and three ablation variants, all running within a single unified
simulation harness.
As discussed in Appendix~\ref{app:related:edu}, existing educational
and agentic systems vary widely in scope and availability; we therefore
design baselines that isolate specific reasoning strategies while
sharing the same backbone (\textit{Gemini-3-Flash}) and KB access as
\textsc{DeepTutor}.
Below we give the algorithmic definition of each baseline, followed by
a brief note on the three ablation variants (w/o SKG, w/o DPM, w/o
SKG + DPM) that selectively disable components of the full pipeline.

\paragraph{Notation.}
All baseline algorithms below share the following primitives.
$\textsc{Rag}(m,\mathcal{K},\text{mode},k)$ retrieves $k$ passages
from knowledge base~$\mathcal{K}$ for query~$m$;
$\textsc{Concat}(\cdot)$ concatenates its arguments into a single
context string; and $\textsc{Llm}(p,\,H,\,\text{ctx})$ denotes one
LLM generation call with system prompt~$p$, conversation history~$H$,
and user-side context~$\text{ctx}$.
Prompt variables $p_{\cdot}$ are symbolic shorthands; their concrete
prompt templates are included in the released repository.

\paragraph{Naive Tutor (\texttt{mock}).}
The simplest baseline: a single LLM call with a minimal tutoring
prompt~$p_{\mathrm{tutor}}$ (instructing the model to act as a helpful
and patient tutor), the conversation history, and a naive-mode
retrieved KB snippet prepended to the current student message.
There is no planning, no tool-execution loop, and no cross-session
memory.

\begin{algorithm}[ht!]
\caption{Naive Tutor (Mock)}\label{alg:mock}
\small
\begin{algorithmic}[1]
\Require Student message $m$, conversation history $H$,
         knowledge base $\mathcal{K}$
\Statex \textbf{Prompt:} $p_{\mathrm{tutor}}$ (tutoring persona)
\State $r_\text{rag} \gets \textsc{Rag}(m,\,\mathcal{K},\,\text{mode}=\text{naive},\,k{=}2)$
\State $\text{ctx} \gets \textsc{Concat}(r_\text{rag},\, m)$
\State $\textit{response} \gets \textsc{Llm}\!\bigl(p_{\mathrm{tutor}},\; H,\; \text{ctx}\bigr)$
\State \Return \textit{response}
\end{algorithmic}
\end{algorithm}

\paragraph{CoT Tutor (\texttt{cot}).}
Identical to the Naive Tutor except that the system prompt
$p_{\mathrm{cot}}$ extends $p_{\mathrm{tutor}}$ with a
chain-of-thought directive~\citep{wei2022chain} (``think step by step
before answering''), asking the model to reason internally before
producing its tutoring response.
The implementation still uses a single generation call.

\begin{algorithm}[ht!]
\caption{CoT Tutor}\label{alg:cot}
\small
\begin{algorithmic}[1]
\Require Student message $m$, conversation history $H$,
         knowledge base $\mathcal{K}$
\Statex \textbf{Prompt:} $p_{\mathrm{cot}}$ (tutoring persona + chain-of-thought directive)
\State $r_\text{rag} \gets \textsc{Rag}(m,\,\mathcal{K},\,\text{mode}=\text{naive},\,k{=}2)$
\State $\text{ctx} \gets \textsc{Concat}(r_\text{rag},\, m)$
\State $\textit{response} \gets \textsc{Llm}\!\bigl(p_{\mathrm{cot}},\; H,\; \text{ctx}\bigr)$
\State \Return \textit{response}
\end{algorithmic}
\end{algorithm}

\paragraph{Self-Refine Tutor (\texttt{self\_refine}).}
Inspired by \citet{madaan2024selfrefine}, this baseline uses two
sequential LLM calls.
The first call uses $p_{\mathrm{tutor}}$ (the same tutoring persona as
\texttt{mock}) to produce an initial draft; the second call uses
$p_{\mathrm{ref}}$, which instructs a pedagogical quality reviewer to
refine the draft for clarity, specificity, and pedagogical
effectiveness.
The refinement pass receives the draft together with the current
student message, conversation history, and the same retrieved context.

\begin{algorithm}[ht!]
\caption{Self-Refine Tutor}\label{alg:selfrefine}
\small
\begin{algorithmic}[1]
\Require Student message $m$, conversation history $H$,
         knowledge base $\mathcal{K}$
\Statex \textbf{Prompts:} $p_{\mathrm{tutor}}$ (tutoring persona),\;
        $p_{\mathrm{ref}}$ (pedagogical reviewer)
\State $r_\text{rag} \gets \textsc{Rag}(m,\,\mathcal{K},\,\text{mode}=\text{naive},\,k{=}2)$
\State $\text{ctx} \gets \textsc{Concat}(r_\text{rag},\, m)$
\State $\textit{draft} \gets \textsc{Llm}\!\bigl(p_{\mathrm{tutor}},\; H,\; \text{ctx}\bigr)$
  \Comment{generate initial response}
\State $\text{ctx}' \gets \textsc{Concat}(r_\text{rag},\, m,\, \textit{draft})$
\State $\textit{response} \gets \textsc{Llm}\!\bigl(p_{\mathrm{ref}},\; H,\; \text{ctx}'\bigr)$
  \Comment{refine for clarity \& pedagogy}
\State \Return \textit{response}
\end{algorithmic}
\end{algorithm}

\paragraph{ReAct Tutor (\texttt{react}).}
This baseline implements a \emph{single-round} ReAct-inspired
loop~\citep{yao2023reactsynergizingreasoningacting} per student turn
via four sequential LLM calls, each with a dedicated prompt:
$p_{\mathrm{think}}$ analyses the student's question and diagnoses
what they most need next;
$p_{\mathrm{act}}$ decides on a concrete tutoring action plan;
$p_{\mathrm{obs}}$ reviews the thought--action pair and notes any
adjustment; and
$p_{\mathrm{tutor}}$ (the shared tutoring persona) synthesises the
accumulated reasoning into a final student-facing response.
All four calls operate over the same retrieved context~$c_0$ and
conversation history.
Relative to \textsc{DeepTutor}, this baseline does not include a
multi-step planner, external tool execution beyond the shared naive
RAG context, or cross-session memory.

\begin{algorithm}[ht!]
\caption{ReAct Tutor (single-round per turn)}\label{alg:react}
\footnotesize
\begin{algorithmic}[1]
\Require Student message $m$, history $H$, knowledge base $\mathcal{K}$
\Statex \textbf{Prompts:} $p_{\text{think}}$, $p_{\text{act}}$, $p_{\text{obs}}$, $p_{\text{tutor}}$
\State $r \gets \textsc{Rag}(m,\mathcal{K},\text{naive},k{=}2)$
\State $c_0 \gets \textsc{Concat}(r, H, m)$ \Comment{base context}
\State $th \gets \textsc{Llm}(p_{\text{think}}, c_0)$ \Comment{diagnose need}
\State $act \gets \textsc{Llm}(p_{\text{act}}, \textsc{Concat}(c_0, th))$
  \Comment{plan action}
\State $obs \gets \textsc{Llm}(p_{\text{obs}}, \textsc{Concat}(c_0, th, act))$
  \Comment{review}
\State $resp \gets \textsc{Llm}(p_{\text{tutor}}, H, \textsc{Concat}(c_0, th, act, obs))$
\State \Return $resp$
\end{algorithmic}
\end{algorithm}

\paragraph{Baseline and ablation metric breakdown.}
Table~\ref{tab:ablation_full} provides the full per-metric breakdown
across the released baselines and \textsc{DeepTutor} variants.

\begin{table*}[ht!]
\centering
\scriptsize
\setlength{\tabcolsep}{4pt}
\resizebox{\textwidth}{!}{%
\begin{tabular}{lcccccccccc}
\toprule
\textbf{Variant}
  & \textbf{SF}$\uparrow$ & \textbf{PER}$\uparrow$
  & \textbf{APP}$\uparrow$ & \textbf{VID}$\uparrow$
  & \textbf{LD}$\uparrow$
  & \textbf{FIT}$\uparrow$ & \textbf{GND}$\uparrow$
  & \textbf{DIV}$\uparrow$ & \textbf{ANS}$\uparrow$
  & \textbf{CC}$\uparrow$ \\
\midrule
\textbf{Naive Tutor}  & 3.2192 & 4.2406 & 4.4438 & 3.8895 & 3.9894 & 3.1978 & 2.4563 & 2.8267 & 3.8744 & 3.1200 \\
\textbf{CoT Tutor}  & 3.3395 & 4.2238 & 4.4709 & 3.8190 & 4.0096 & 3.1800 & 2.4978 & 2.7911 & 3.8120 & 3.0444 \\
\textbf{Self-Refine Tutor}  & 3.2785 & 4.2756 & \textbf{4.6090} & 3.9384 & 4.1414 & 3.1711 & 2.4911 & 2.8756 & 3.8844 & 2.9689 \\
\textbf{ReAct Tutor}  & 3.3472 & 4.3722 & 4.4023 & 3.7044 & 3.9598 & 3.1578 & 2.4667 & 2.7511 & \underline{3.9444} & 3.0667 \\
\midrule
\textsc{DeepTutor}  & \textbf{3.3641} & \textbf{4.5930} & 4.5638 & 4.8145 & \textbf{4.6079} & \textbf{3.3500} & \textbf{2.9600} & \textbf{3.4422} & \textbf{3.9767} & \textbf{3.3822} \\
w/o SKG    & 3.1057 & \underline{4.4091} & 4.5537 & \textbf{4.8316} & \underline{4.5348} & \underline{3.3055} & 2.2311 & 3.3022 & 3.8355 & 3.1911 \\
w/o DPM & \underline{3.3554} & 4.2243 & 4.5695 & \underline{4.8269} & 4.5156 & 3.1538 & \underline{2.8022} & \underline{3.3400} & 3.9011 & \underline{3.3089} \\
w/o SKG + DPM  & 3.1609 & 4.1724 & \underline{4.5983} & 4.8200 & 4.4811 & 3.2533 & 2.2211 & 3.2733 & 3.8522 & 3.1356 \\
\bottomrule
\end{tabular}
}
\caption{Full metric breakdown across the released baselines and
  \textsc{DeepTutor} variants.
  All columns are 1--5 averages, filled from the released overall
  evaluation summary manifest.
  Bold indicates the best value in each column; underline indicates the
  second-best value.
  Abbreviations: SF~=~source faithfulness; PER~=~personalization;
  APP~=~applicability; VID~=~vividness; LD~=~logical depth;
  FIT~=~fitness; GND~=~groundedness; DIV~=~diversity;
  ANS~=~answer quality; CC~=~cross concept.}
\label{tab:ablation_full}
\end{table*}

% ------------------------------------------------------------
\subsection{Human Preference Alignment Study}
\label{app:eval:human_alignment}
% ------------------------------------------------------------

The human alignment study in \S\ref{sec:human_alignment} is designed to
test whether the LLM judge's relative system preferences agree with
independent human judgments, rather than to measure downstream learning
gains.

\paragraph{Sampling and materials.}
We use a domain-stratified subset of 45 TutorBench sessions, with nine
sessions sampled from each of the five domains used in
Table~\ref{tab:domain}. For every selected session, we collect the full
\textsc{DeepTutor} and \textit{Naive Tutor (Mock)} outputs generated
under the same benchmark entry: identical learner profile, task
description, source material, student-simulator protocol, and
practice-generation request. The comparison packet shown to annotators
contains the learner profile, task description, source material, tutoring
transcript, and generated practice questions for both systems.

\paragraph{Human annotation.}
We recruit ten student annotators with relevant academic training: six
undergraduate students and four graduate students (two master's students
and two Ph.D. students). Each domain is assigned to two annotators.
Research sessions are evaluated by the two Ph.D. students; Engineering
and Business each include one domain-related master's student; the
remaining domain assignments are completed by domain-related
undergraduates. Each packet is evaluated by both assigned annotators in
a blind A/B interface.
System names are hidden, and the order of A and B is randomly flipped
per session to reduce order bias. Annotators are given the same ten
metric definitions used by the LLM judge: Source Faithfulness,
Personalization, Applicability, Vividness, Logical Depth, Fitness,
Groundedness, Diversity, Answer Quality, and Cross Concept. For each
metric, the annotator selects one of three labels: A is better, B is
better, or the two systems are indistinguishable. This pairwise design
avoids the scale-calibration problem of absolute 1--5 human scoring while
directly testing the system preference implied by the automatic judge.

\paragraph{Aggregation and LLM comparison.}
Human labels are first mapped back to system identities and then
aggregated for each session--metric pair; when the two annotators do not
select the same system preference, the aggregate label is treated as a
tie. For the LLM side, we use the
corresponding LLM-judge preference label for the same session and
metric. Figure~\ref{fig:human_alignment} reports, for each metric, the
share of comparisons where humans prefer \textsc{DeepTutor}, mark a tie,
or prefer Mock, alongside the corresponding LLM-judge preference shares.

\paragraph{Alignment statistics.}
The correlation statistics reported in
Figure~\ref{fig:preference_align_distribution} and
\S\ref{sec:human_alignment} are computed over the ten metric-level
\textsc{DeepTutor} win-rate pairs, rather than over all individual
session--metric labels. Pearson $r=0.82$ ($p=0.0038$) and Spearman
$\rho=0.83$ ($p=0.0027$) therefore test whether the rubric dimensions most favored by
human raters are also the dimensions most favored by the LLM judge. At
the same metric-summary level, the directional winner agrees on all ten
metrics: both human raters and the LLM judge assign the largest
preference share to \textsc{DeepTutor} for every metric.

% ------------------------------------------------------------
\subsection{General Problem-Solving Setup}
\label{app:eval:general}
% ------------------------------------------------------------

\subsubsection{Benchmark Descriptions}

Table~\ref{tab:benchmarks} summarizes the five public benchmarks used
in \S\ref{sec:generalization}.

\begin{table*}[t]
\centering
\small
\setlength{\tabcolsep}{6pt}
\resizebox{\textwidth}{!}{%
\begin{tabular}{llccl}
\toprule
\textbf{Benchmark} & \textbf{Category} & \textbf{Size} & \textbf{Subset} & \textbf{Format} \\
\midrule
HLE~\citep{phan2025hle}
  & STEM reasoning & $\sim$2,500 & 500 (text+mm) & open-ended \\
GPQA-Diamond~\citep{rein2024gpqa}
  & STEM reasoning & 198 & all 198 & 4-choice MCQ \\
LiveBench~\citep{white2025livebench}
  & reasoning & $\sim$200 & reasoning cat. & mixed \\
GAIA~\citep{mialon2024gaia}
  & agentic & $\sim$165 & validation (L1--L3) & open-ended \\
AA-LCR~\citep{artificialanalysis2025lcr}
  & long-context & 100 & all 100 & open-ended \\
\bottomrule
\end{tabular}
}
\caption{Public benchmarks used in the general problem-solving evaluation.
  ``Size'' refers to the total available test items; ``Subset'' is the
  number we evaluate on. mm~=~multimodal (text + images).}
\label{tab:benchmarks}
\end{table*}

\subsubsection{Evaluation Protocol}

Correctness is assessed via a two-stage pipeline.
\textbf{Stage~1 (Extract)}: an LLM extracts the model's final answer
from its free-form output (model: Claude~Sonnet~4.6, temperature~=~0.0,
max tokens~=~16384).
\textbf{Stage~2 (Judge)}: a second LLM call compares the extracted
answer to the ground truth using benchmark-specific rubrics (same
model and settings as Stage~1 for consistency).
Benchmark-specific rubrics cover exact match for GAIA, option matching or text equivalence for GPQA and HLE, and textual equivalence checks for open-ended benchmarks such as LiveBench and AA-LCR.
All Pass@1 scores are reported as percentages.

\subsubsection{Pipeline Configuration}

Table~\ref{tab:pipeline_config} lists the tool configuration and
concurrency settings for the \textsc{DeepTutor} pipeline on each
benchmark.
This is the same solver-only configuration described in
\S\ref{sec:generalization}: the Hybrid Personalization Engine is
disabled, including both Static Knowledge Grounding (SKG) and Dynamic
Personal Memory (DPM).
Thus no course knowledge base, learner profile, or trace-forest
retrieval is active in the public benchmark runs.

\begin{table}[ht!]
\centering
\scriptsize
\setlength{\tabcolsep}{4pt}
\resizebox{\columnwidth}{!}{%
\begin{tabular}{lcccc}
\toprule
\textbf{Benchmark} & \textbf{Tools Enabled} & \textbf{Max Tokens} & \textbf{Concurrency} \\
\midrule
HLE         & code\_execute, reason   & 16384 & 30 \\
GPQA-D      & code\_execute, reason   & 16384 & 50 \\
LiveBench   & code\_execute, reason   & 16384 & 10 \\
GAIA        & code\_execute, reason, web\_search & 16384 & 20 \\
AA-LCR      & code\_execute, reason   & 16384 & 25 \\
\bottomrule
\end{tabular}
}
\caption{Per-benchmark pipeline configuration.
  \texttt{done} and \texttt{replan} are always available.
  Temperature is 0.0 across all benchmarks.}
\label{tab:pipeline_config}
\end{table}

% ------------------------------------------------------------
\subsection{Agent Hyperparameters}
\label{app:eval:hparams}
% ------------------------------------------------------------

Table~\ref{tab:hparams} reports the temperature and maximum token
budget for each module.
The solve module uses low temperature (0.3) for deterministic
multi-step reasoning; the question module uses higher temperature (0.7)
to promote ideation diversity; the personalization module uses
an intermediate value (0.5) balancing consistency with adaptive
profile updates.

\begin{table}[ht!]
\centering
\scriptsize
\setlength{\tabcolsep}{4pt}
\resizebox{\columnwidth}{!}{%
\begin{tabular}{llcc}
\toprule
\textbf{Module} & \textbf{Agents} & \textbf{Temp.} & \textbf{Max Tokens} \\
\midrule
Solve
  & planner, solver, writer & 0.3 & 8192 \\
Question
  & idea, evaluator, generator, validator & 0.7 & 4096 \\
Personalization
  & summary, weakness, reflection & 0.5 & 8192 \\
\bottomrule
\end{tabular}
}
\caption{Agent hyperparameters.
  All modules share a single set of parameters; per-agent overrides
  are not used in the reported experiments.}
\label{tab:hparams}
\end{table}

% ------------------------------------------------------------
\subsection{Implementation Details}
\label{app:eval:impl}
% ------------------------------------------------------------

For reproducibility, we detail key implementation choices not fully
covered in the main text.

\paragraph{Weakness State Transitions.}
Weakness state transitions (active $\to$ resolved) are
\emph{evidence-gated} rather than governed by numeric confidence
thresholds.
The Weakness Agent uses prompt-based criteria: a weakness is marked
resolved when the user answers correctly on the topic in recent
sessions \emph{and} stops re-asking about it.
The benchmark evaluator enforces strict resolution criteria---explicit
correction of the misconception, a clear explanation of why the old
understanding was wrong, and scaffolded reasoning for complex concepts.
Profile evolution moves resolved concepts into \texttt{known\_well} and
removes them from \texttt{unknown} and \texttt{partially\_known}.

\paragraph{Trace Retrieval Budget.}
Trace retrieval uses a fixed \texttt{top\_k}~=~3 for semantic search
across the trace forest.
Level-specific filtering is applied by context: idea and evaluator
agents query level-1 nodes (session roots); generator context queries
level-3 nodes for wrong-answer patterns.
The \textsc{SearchTrace} tool defaults to \texttt{top\_k}~=~5, and
individual trace detail views are truncated to 6{,}000 characters.
Scratchpad context budgets are 6{,}000 tokens for the solver and
12{,}000 tokens for the writer.
Conversation history in the benchmark simulator is truncated to
10{,}000 characters.

\paragraph{Embedding Models.}
Both the trace forest and RAG indexing share the same embedding service.
The embedding model is configured via environment variables;
supported options include OpenAI \texttt{text-embedding-3-large}
(3{,}072 dimensions), Cohere \texttt{embed-v4.0} (1{,}024 dimensions),
and Jina \texttt{jina-embeddings-v3} (1{,}024 dimensions).

\paragraph{Knowledge Graph Construction.}
Knowledge graph construction is delegated to
LightRAG, which performs chunking, entity extraction, and embedding
internally.
RAGAnything adds MinerU-based PDF parsing (images, tables, equations)
before insertion.
Our implementation uses LightRAG's default entity extraction pipeline
without custom entity alignment or additional quality-control filtering.

\paragraph{Additional Solver Parameters.}
Table~\ref{tab:solver_params} lists additional parameters governing the
solver-only and problem-tutoring pipelines.

\begin{table}[ht!]
\centering
\scriptsize
\setlength{\tabcolsep}{3pt}
\begin{tabular*}{\columnwidth}{@{\extracolsep{\fill}} l c l c}
\toprule
\textbf{Parameter} & \textbf{Value} & \textbf{Parameter} & \textbf{Value} \\
\midrule
Max ReAct iterations per step & 5 & RAG mode (planner) & hybrid \\
Max plan steps & 10 & Question idea loop rounds & 3 \\
Max replans & 2 & Ideas per round & 5 \\
Observation truncation & 2{,}000 tokens & Generator max retries & 2 \\
Planner pre-retrieval queries & 3 & Question RAG mode & naive \\
RAG \texttt{top\_k} (planner) & 8 & Memory agent max ReAct rounds & 6 \\
\bottomrule
\end{tabular*}
\caption{Additional pipeline parameters.}
\label{tab:solver_params}
\end{table}

% ------------------------------------------------------------
\subsection{Backbone Models}
\label{app:eval:backbones}
% ------------------------------------------------------------

Table~\ref{tab:backbones} lists the five backbone families evaluated
in the general problem-solving experiments (\S\ref{sec:generalization}).
All backbone results are obtained via OpenRouter-compatible APIs with
greedy decoding (temperature~=~0.0).

\begin{table}[ht!]
\centering
\small
\begin{tabular}{lll}
\toprule
\textbf{Backbone} & \textbf{Provider} & \textbf{Context Window} \\
\midrule
Gemini-3-Flash  & Google DeepMind & 1M tokens \\
Sonnet-4.5      & Anthropic        & 200K tokens \\
Qwen-3.5-Plus   & Alibaba Cloud    & 128K tokens \\
GPT-5-Mini      & OpenAI           & 128K tokens \\
Minimax-M2.5    & MiniMax          & 1M tokens \\
\bottomrule
\end{tabular}
\caption{Backbone families used in the general problem-solving evaluation.
  Context window sizes are approximate as reported at evaluation time.}
\label{tab:backbones}
\end{table}

% ============================================================

\end{document}